\documentclass[journal]{IEEEtran}
\usepackage{float}
\usepackage{graphicx} %package to manage images
\usepackage{amsmath}
\usepackage{color}
\usepackage{amsfonts,amssymb}
\usepackage{booktabs}
\usepackage{multirow} 

\ifCLASSINFOpdf
\else
\fi
\begin{document}
\title{Resilient Secondary Voltage Control of Islanded Microgrids: An ESKBF-Based Distributed Fast Terminal Sliding Mode Control Approach}

\author{Pudong~Ge,~\IEEEmembership{Student Member,~IEEE,}
        Yue~Zhu,
       Tim~C.~Green,~\IEEEmembership{Fellow,~IEEE} and~Fei~Teng,~\IEEEmembership{Member,~IEEE}% <-this % stops a space

\thanks { This work was supported by EPSRC under Grant EP/T021780/1 and by ESRC under Grant ES/T000112/1.
}

\thanks {The authors are with the Department of Electrical and Electronic Engineering, Imperial College London, London SW7 2AZ, U.K.  Corresponding author:  Dr  Fei  Teng (f.teng@imperial.ac.uk).} % <-this % stops a space
}

\maketitle

\begin{abstract}
This paper proposes a distributed secondary voltage control method based on extended state Kalman-Bucy filter (ESKBF) and fast terminal sliding mode (FTSM) control for the resilient operation of an islanded microgrid (MG) with inverter-based distributed generations (DGs). 
%The existing distributed algorithms are commonly designed considering one type of uncertainty in the MG system and cannot fully investigate the complex uncertainties. However, different types of uncertainties exist in the MG and probably affect the MG robustness together. Therefore, in this paper,
To tackle the co-existence of multiple uncertainties, a unified modelling framework is proposed to represent the set of different types of disturbances, including parameter perturbation, measurement noise, and immeasurably external variables, by an extended state method. Kalman-Bucy filter is then applied to accurately estimate the state information of the extended DG model. %that can be significantly influenced by model uncertainties. 
In addition, based on the accurate estimation, a fast terminal sliding mode (FTSM) surface with terminal attractors is designed to maintain the system stability and accelerate the convergence of consensus tracking, which significantly improves the performance of secondary voltage control under both normal and plug-and-play operation. Finally, case studies are conducted in both MATLAB/Simulink and an experimental testbed to demonstrate the effectiveness of the proposed method.
\end{abstract}


% \fi
\IEEEpeerreviewmaketitle

\section{Introduction}
\IEEEPARstart{M}{icrogrid} (MG) is a promising concept that supports the integration of distributed generation (DG) into traditional bulk electric power systems ~\cite{olivares_trends_2014,antoniadou-plytaria_distributed_2017}. 
A MG is a group of interconnected loads and distributed energy resources within clearly defined electrical boundaries that acts as a single controllable entity with respect to the grid, and it can connect and disconnect from the grid to enable it to operate in both grid-connected and islanded (autonomous) modes ~\cite{ton_u.s._2012}.
%A MG typically consists of renewable, non-renewable generators and energy storage units, and can operate in both grid-connected and islanded (autonomous) modes~\cite{mehrizi-sani_potential-function_2010}. 
While the dynamics of grid-connected MG are determined by the main grid, the stability of islanded MG highly relies on the underlining control strategy~\cite{olivares_trends_2014,farrokhabadi_microgrid_2019}.

To stabilize islanded MGs, a three-layer control structure has been proposed, including primary, secondary and tertiary control~\cite{wu_review_2018,yu_control_2010,guerrero_hierarchical_2011}. Primary control~\cite{wu_review_2018} is implemented locally to guarantee the stability of MGs by only using the local DG's information~\cite{diaz_scheduling_2010}, while at the same time, secondary and tertiary control~\cite{guerrero_hierarchical_2011,zhang_three-stage_2019,li_dynamic_2017} are employed to ensure the voltage and frequency of MG being regulated to the references and the optimal power sharing. In general, the different levels differ in control goals and infrastructure requirements (i.e.,communication requirements)~\cite{olivares_trends_2014}. The microgrid can benefit from this control hierarchy by decoupling control goals in terms of response speed and time frame, and DGs can respond to the load change independently thanks to the primary control that does not require communication.

The conventional secondary control is designed in a centralized mode~\cite{diaz_centralized_2017,zhong_robust_2013}, where DGs receive control commands from a center controller. Due to the reliance on the communication network, the key drawbacks of centralized approach are: 1) high communication delays that might degrade the control performance; 2) poor plug-and-play capability; 3) low fault tolerance due to all-to-one control scheme~\cite{olivares_trends_2014,lewis_secondary_2013}.% low reliability and sensitive to a single point of failure.

These drawbacks have been overcome by the development of distributed control scheme~\cite{cady_distributed_2015,etemadi_decentralized_2012} and multi-agent system~\cite{lewis_secondary_2013,dou_multiagent_2017}. %A combination of distributed control and multi-agent theory is widely undertaken in the MG control due to its extensibility, scalability, flexibility and communication reliability~\cite{bidram_distributed_2013,zhang_distributed_2017,wu_distributed_2018}. 
In the existing literature for the distributed control of MG, the feedback linearization has been widely used to simplify the modelling of DG inner control loops by transferring the highly-nonlinear MG voltage control model into a single-input single-output second-order model~\cite{bidram_distributed_2013,zuo_distributed_2016}. Nevertheless, after the application of feedback linearization, one complex variable is required to represent the total non-linearity, consisting of many voltage and current variables. 
This complex variable is contaminated by inevitable measurement noise, the effect of which on the linearized MG control model has not been fully investigated. Thus, it is vital to evaluate and potentially eliminate the influence of measurement noise in the control system design.
%%model simplification methods \textcolor{red}{(to be explained)}~\cite{rasheduzzaman_reduced-order_2015} are still under developing. To fully consider the impact of DG inner control loops, feedback linearization has been widely used and it can transfer the detailed MG voltage control model into a single-input single-output second-order model~\cite{bidram_distributed_2013,zuo_distributed_2016}, which is friendly to control strategy design. 

Moreover, the co-existence of disturbances caused by different sources such as parameter perturbation and measurement noise has not been fully investigated. Although robust distributed control~\cite{etemadi_generalized_2014,islam_distributed_2017} and noise-resilient control~\cite{dehkordi_distributed_2019,ge_extended-state-observer-based_2020} have been applied respectively, these control methods cannot simultaneously consider multiple types of disturbances in an efficient manner. The effect of the measurement noise on the MG secondary control has not been considered in~\cite{bidram_distributed_2013,zuo_distributed_2016,dehkordi_distributed_2017,pilloni_robust_2018}, though these methods show good robustness against parameter perturbation or other unmodeled uncertainty.
The existing methods tend to consider the multiple disturbances separately, while a systematic and unified disturbance modelling framework has yet to be developed. %for the MG control system. 
The recent development of advanced disturbance-resilient methods in control theory, such as adaptive filter algorithms~\cite{sun_distributed_2018,zhang_extended_2018}, provides a strong theoretical basis to develop such modelling framework.

In addition, plug-and-play operation in the multi-DG network may lead to frequent and significant voltage fluctuations, which imposes a vast challenge on the fast restoration of the MG voltage. In this circumstance, the convergence rate in the distributed control of MG has received increasing attention. Some distributed control methods~\cite{zuo_distributed_2016,dehkordi_distributed_2017,pilloni_robust_2018} has been designed for MGs to achieve finite-time convergence rate. However, the convergence of finite-time control protocol relies on its initial states, and the plug-and-play operation may cause unknown or diverse system initial states. For the MG system that emphasizes on the plug-and-play capability, a novel control strategy for resilience enhancement is required to alleviate such impact by improving the convergence performance.

To mitigate the aforementioned problems, a distributed robust fast terminal sliding mode (FTSM) secondary voltage control method based on extended state Kalman-Bucy filter (ESKBF) is proposed in this paper. The ESKBF employs an extended state to denote the combination of different types of uncertainties, including parameter perturbation, measurement noise and immeasurably external variables. The proposed FTSM control enhances the convergence rate of MG voltage control, where the settling time of the distributed controller can be reduced. The main features and contributions of the proposed control method is summarized as follows:
\begin{enumerate}
    \item Linearized control-oriented model formulation under a unified modelling framework for multiple sources of disturbances: 
    %this paper focuses on the uncertainties in the MG control model caused by parameter perturbation, measurement noise and immeasurable variables uniformly in the control-oriented model. 
    unlike the traditional disturbance observer that targets at filtering out exact disturbance magnitude, %a particular type of disturbance, 
    we integrate the multiple disturbances related to parameter perturbation, immeasurable variables and measurement noise into one extended state which was originally used to only represent the complicated nonlinear part of MG model caused by feedback linearization. %In addition, the measurement noise is represented by additive variables in the control-oriented MG model.
    This formulation significantly simplifies the observer design under multiple sources of disturbance.
    \item Multi-disturbance resilient observer design: to obtain the accurate model state for the control implementation, the impact of multiple-source disturbance is first described by a combination of process noise and observation noise in the system model. Kalman-Bucy filter is then utilized to design a multi-disturbance resilient observer that can estimate the extended state value and cope with stochastic measurement noise without complicated parameter selection process required in the existing extended state observer design~\cite{madonski_survey_2015}.
    %\textcolor{red}{no parameter needs to be selected?}
    \item Faster convergence rate: to accelerate the convergence rate in the MG distributed voltage control for plug-and-play operation, a FTSM surface is designed by employing nonlinear terminal attractors to guarantee a faster convergence rate when the system is close to equilibrium. To achieve this, we propose a nonlinear control protocol and prove its global stability. Through the coordination of FTSM control and the proposed nonlinear control protocol, the voltage control of MG can achieve globally consensus stability with short settling time. In addition, this control framework can be extended to balance voltage regulation and accurate reactive power sharing.
\end {enumerate}

This paper is structured as follows. In Section~\ref{section_pre}, preliminary notions of graph theory and the detailed model of islanded MG are introduced. Section~\ref{section_eskbf} introduces the ESKBF-based observer for MG voltage control, and in Section~\ref{section_vol_ctrl}, the FTSM secondary voltage control is discussed. Then, Section~\ref{section_implement} illustrates the implementation structure of the proposed control scheme. Finally, simulation and experimental results are analyzed in Section~\ref{section_simulation} and the conclusion and future work are discussed in Section~\ref{section_conclusion} .

\section{Preliminaries and Model Description}\label{section_pre}
\subsection{Preliminary of Graph Theory}
The communication topology among DGs in a MG can be modeled as a weighted graph $\mathcal{G=\{V,E,A}\}$ with a node set $\mathcal{V}=\{v_1,v_2,\cdots,v_N\}$, an edge set $\mathcal{E}\subset\mathcal{V}\times\mathcal{V}$, and a weighted adjacent matrix $\mathcal{A}=[a_{ij}]\in\mathbb{R}^{N\times N}$. If DG node $v_i$ can receive information from DG node $v_j$, edge $(v_j,v_i)\in\mathcal{E}$ and set $N_i=\{j|(v_j,v_i)\in\mathcal{E}\}$ means neighbors of node $i$. For adjacent matrix $\mathcal{A}$, elements $a_{ii}=0$ and $a_{ij}\ge0$. $a_{ij}>0$ if and only if $(v_i,v_j)\in\mathcal{E}$. The Laplacian matrix of $\mathcal{G}$ is defined as $\mathcal{L}=[l_{ij}]=\mathcal{D-A}\in\mathbb{R}^{N\times N}$, where $\mathcal{D}={\rm diag}\{d_i\}$ denotes the in-degree matrix with $d_i=\sum_{j\in N_i}{a_{ij}}$~\cite{qu_cooperative_2009}.

A MG can be modeled as a leader-following multi-agent system with $N$ DGs. In this leader-following structure, the frequency and voltage references are only available to a small portion of DGs (as ``leader"). Other DGs (as ``followers") have neighboring information through a sparse communication structure, and then ``followers" can track to the ``leader" DGs' frequency and voltage.
The adjacency matrix extended by the reference node is denoted as $\mathcal{B}={\rm diag}\{b_i\}\in\mathbb{R}^{N\times N}$, in which if the reference node is a neighbor of node $v_i$, $b_i>0$; otherwise, $b_i=0$. The corresponding Laplacian matrix $\overline{\mathcal{L}} = \mathcal{L+B}$.

\subsection{Large-Signal Dynamic Model of MGs with Inverter-Based DGs}

\begin{figure}[!t]
\centering
\includegraphics[width=0.48\textwidth]{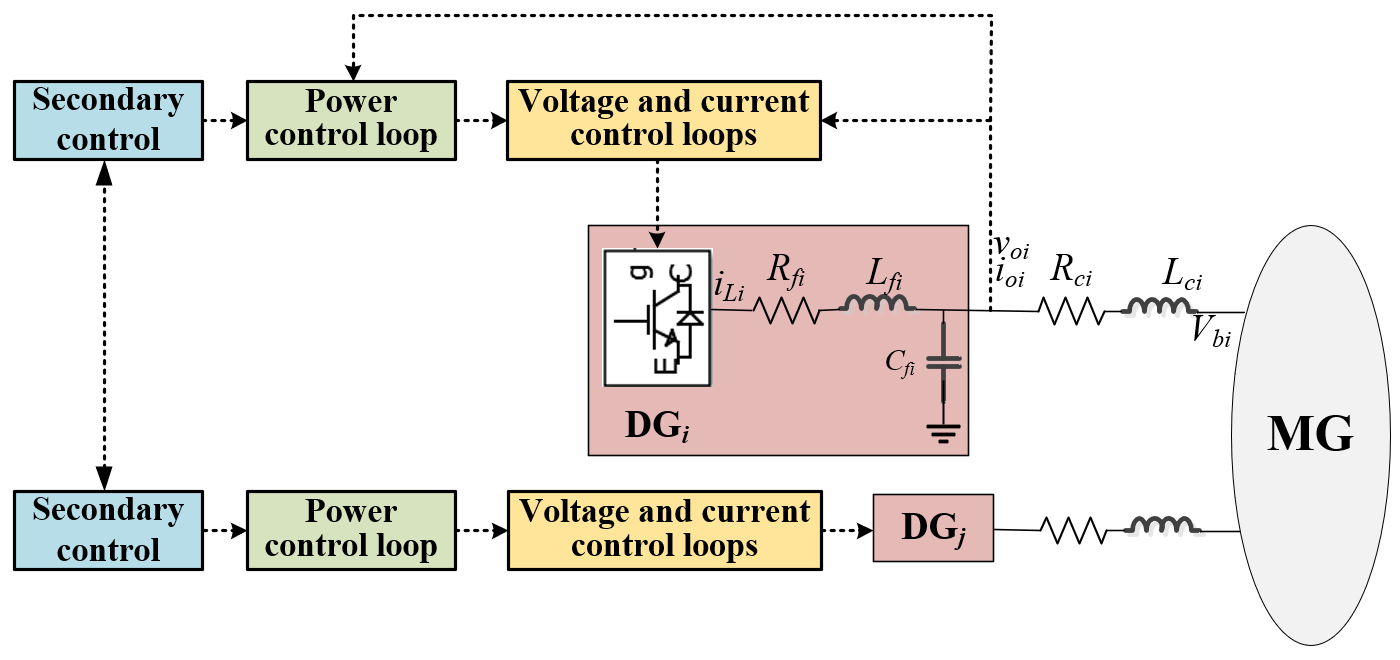}
\caption{Block diagram of an inverter-based MG.}
\label{fig_MG}
\end{figure}

As depicted in Fig.\ref{fig_MG}, each DG unit contains a DC/AC inverter, an inductor-capacitor (LC) filter and a resistor-inductor connection, and its controller is composed of three control loops formulated on its own direct-quadrature frame $(d-q)_i$ at rotating frequency $\omega_i$: power control loop, voltage control loop and current control loop~\cite{bidram_distributed_2013,zuo_distributed_2016}.

The power control loop of each DG controller generates the angular frequency and voltage references for the whole DG system, while voltage and current loops enable the inverter system to track the angular frequency and voltage references. Droop control~\cite{diaz_scheduling_2010} is typically used in the power control loop, and it can describe the relationship between angular frequency, output voltage magnitude and power output:
\begin{align}
     \omega_i &= \omega_{ni}-m_{Pi}P_i\\
     v_{o,magi}^* &= v_{odi}^* = V_{ni}-n_{Qi}Q_i\\
     v_{oqi}^* &= 0
\end{align}
where $\omega_i$ is the angular frequency of the $i$th DG; $v_{odi}^*$ and $v_{oqi}^*$ are the $d$-axis and $q$-axis components of $v_{o,magi}^*$, and $v_{odi}^*=v_{o,magi}^*$ due to the fact that $v_{o,magi}^*$ often aligns itself on the $d$-axis; $P_i$ and $Q_i$ are the active power and reactive power; $m_{Pi}$ and $n_{Qi}$ are the frequency and voltage droop coefficients, which are selected based on the active and reactive power ratings of each DG~\cite{guerrero_hierarchical_2011}; $\omega_{ni}$ and $V_{ni}$ are references for primary control that are generated from the secondary control.

The power elements $P_i$ and $Q_i$ are obtained by instantaneous power calculation and low-pass filter, which can be expressed as follows:
\begin{align}
     P_i = \frac{\omega_{ci}}{s+\omega_{ci}}(v_{odi}i_{odi}+v_{oqi}i_{oqi})\\
     Q_i = \frac{\omega_{ci}}{s+\omega_{ci}}(v_{oqi}i_{odi}-v_{odi}i_{oqi})
\end{align}
where $v_{odi}, v_{oqi}, i_{odi}, i_{oqi}$ are the $d$-axis and $q$-axis components of output voltage $v_{oi}$ and output current $i_{oi}$. $\omega_{ci}$ is the cut-off frequency of low-pass filter.

To model all DG inverters synchronized in the common frequency frame with the rotating frequency $\omega_{com}$, $\delta_i$ is employed to represent the angular frequency difference of the $i$th DG unit compared to the common reference frame:
\begin{align}
    \dot{\delta} _i=\omega_i-\omega_{com}
\end{align}

The dynamics of voltage and current control loops have been discussed in~\cite{bidram_distributed_2013,zuo_distributed_2016,pilloni_robust_2018}, and are omitted here. By combining dynamic models of the three control loops and LC filter, the large-signal dynamic model of the $i$th DG over a MG system can be detailed as the following multi-input multi-output (MIMO) nonlinear system:
\begin{align}
    \boldsymbol{\dot{x}}_i = \boldsymbol{f}_i(\boldsymbol{x}_i)+\boldsymbol{g}_i(\boldsymbol{x}_i)\boldsymbol{u}_i+\boldsymbol{k}_i(\boldsymbol{x}_i)\boldsymbol{d}_i\label{eq_MG_Model}
\end{align}
where the state vector is 
\begin{align*}
    \boldsymbol{x}_i = [\delta_i\ P_i\ Q_i\ \phi_{di}\ \phi_{qi}\ \gamma_{di}\ \gamma_{qi}\ i_{ldi}\ i_{lqi}\ v_{odi}\ v_{oqi}\ i_{odi}\ i_{oqi}]^T
\end{align*}
and system input $\boldsymbol{u}_i=[\omega_i\ V_{ni}]^T$; $\boldsymbol{d}_i=[\omega_{com}\ v_{bdi}\ v_{bqi}]^T$ are the considered disturbance.

It should be noted that the dispatchable DGs considered in this paper are controlled by grid-forming inverters~\cite{rocabert_control_2012}. A class of grid-forming inverters with the droop control loop have been widely adopted in the current secondary control scheme of MGs~\cite{dehkordi_distributed_2017,pilloni_robust_2018,zuo_distributed_2016_1}. The DGs that operate in maximum power tracking mode are normally controlled by grid-supporting or grid-feeding inverters, and dispatched in different layers~\cite{bidram_multiobjective_2014}.

\section{Extended State Observer Design for Secondary Voltage Control}\label{section_eskbf}

In this section, to enhance the resilience and robustness of secondary voltage control of MG, an extended state observer is designed to mitigate the influence of multiple sources of disturbance on the control performance.

\subsection{Model Linearization of Nonlinear Systems}
For the secondary voltage control of MG, the $i$th DG system model (\ref{eq_MG_Model}) can be described as the following single-input single-output (SISO) system:
\begin{align}
    \left\{
    \begin{aligned}
        \boldsymbol{\dot{x}}_i &= \boldsymbol{f}_i(\boldsymbol{x}_i)+\boldsymbol{g}_i(\boldsymbol{x}_i){u}_i+\boldsymbol{k}_i(\boldsymbol{x}_i)\boldsymbol{d}_i\\
        &= \boldsymbol{F}_i(\boldsymbol{x}_i)+\boldsymbol{g}_i(\boldsymbol{x}_i){u}_i\\
        y_i &= v_{odi} = h_i(\boldsymbol{x}_i)
    \end{aligned}
    \right.\label{eq_Nonlinear}
\end{align}
where $u_i=V_{ni}$.

By applying feedback linearization method, the nonlinear system (\ref{eq_Nonlinear}) can be transformed into:
\begin{align}
\left\{
    \begin{aligned}
        \dot{y}_{i,1} &= \dot{v}_{odi} = y_{i,2}\\
        \dot{y}_{i,2} &= \ddot{v}_{odi} = L_{\boldsymbol{F}_i}^{2}h_i(\boldsymbol{x}_i)+L_{\boldsymbol{g}_i}L_{\boldsymbol{F}_i}h_i(\boldsymbol{x}_i)u_i\\
        y_i &= y_{i,1} = v_{odi}
    \end{aligned}
    \right.\label{eq_Linear}
\end{align}
and
\begin{align}
\left\{
    \begin{aligned}
        L_{\boldsymbol{F}_i}h_i(\boldsymbol{x}_i) &= \frac{\partial h_i}{\partial \boldsymbol{x}_i}\boldsymbol{F}_i(\boldsymbol{x}_i)\\
        L_{\boldsymbol{F}_i}^{k}h_i(\boldsymbol{x}_i) &= \frac{\partial L_{\boldsymbol{F}_i}^{k-1}h_i}{\partial \boldsymbol{x}_i}\boldsymbol{F}_i(\boldsymbol{x}_i)\\
        L_{\boldsymbol{g}_i}L_{\boldsymbol{F}_i}^{k-1}h_i(\boldsymbol{x}_i) &= \frac{\partial L_{\boldsymbol{F}_i}^{k-1}h_i}{\partial \boldsymbol{x}_i}\boldsymbol{g}_i(\boldsymbol{x}_i)
    \end{aligned}
    \right.\label{eq_Lie}
\end{align}
where $L_{\boldsymbol{F}_i}h_i(\boldsymbol{x}_i)$ represents the Lie derivative~\cite{khalil_nonlinear_2001} of $h_i(\boldsymbol{x}_i)$ along ${\boldsymbol{F}_i}(\boldsymbol{x}_i)$. 

Define $z_i=L_{\boldsymbol{F}_i}^{2}h_i(\boldsymbol{x}_i)+L_{\boldsymbol{g}_i}L_{\boldsymbol{F}_i}h_i(\boldsymbol{x}_i)u_i$, the system (\ref{eq_Linear}) can be expressed as a second-order linear system:
\begin{align}
\left\{
    \begin{aligned}
        \dot{y}_{i,1} &= y_{i,2}\\
        \dot{y}_{i,2} &= z_i\\
        y_i &= y_{i,1} = v_{odi}
    \end{aligned}
    \right.\label{eq_sec_ord}
\end{align}
and the input of the nonlinear system (\ref{eq_Nonlinear}) is
\begin{align}
    u_i = \frac{z_i-L_{\boldsymbol{F}_i}^{2}h_i(\boldsymbol{x}_i)}{L_{\boldsymbol{g}_i}L_{\boldsymbol{F}_i}h_i(\boldsymbol{x}_i)}\label{eq_input_trans}
\end{align}

\subsection{Control-Oriented Model Formulation Considering System Disturbances}\label{subsec_control-oriented model}
Although the second-order linear system model (\ref{eq_sec_ord}) is simple and convenient for the design of distributed secondary voltage control, it is difficult to obtain the accurate value of system  state $y_{i,2}$ in (\ref{eq_sec_ord}) and the original system input $u_i$ in (\ref{eq_input_trans}). 
%%which are vital to the whole control system. 

The state $y_{i,2}=\dot{v}_{odi}$ is in the differential form and can not be directly obtained in the industrial practice. Therefore, to avoid the differential operation, the state can be accessed by the equation described in (\ref{eq_MG_Model}):
\begin{align}
    \dot{v}_{odi} = {\omega_i}{v}_{oqi}+\frac{i_{ldi}-i_{odi}}{C_{fi}} \label{eq_dv}
\end{align}

To obtain the original input $u_i$ in (\ref{eq_input_trans}), the two variables, $L_{\boldsymbol{F}_i}^{2}h_i(\boldsymbol{x}_i)$ and $L_{\boldsymbol{g}_i}L_{\boldsymbol{F}_i}h_i(\boldsymbol{x}_i)$, which consist of a large number of measurement variables, can be expressed as
\begin{align}
   \begin{aligned}
        L_{\boldsymbol{F}_i}^{2}h_i(\boldsymbol{x}_i) = & (-\omega_{i}^{2}-\frac{K_{Pci}K_{Pvi}+1}{C_{fi}L_{fi}}-\frac{1}{C_{fi}L_{ci}})v_{odi}\\
        & -\frac{\omega_b{K_{Pci}}}{L_{fi}}{v_{oqi}}+\frac{R_{ci}}{C_{fi}L_{ci}}i_{odi}-\frac{2\omega_i}{C_{fi}}i_{oqi}\\
        & -\frac{R_{fi}+K_{Pci}}{C_{fi}L_{fi}}i_{ldi}+\frac{2\omega_i-\omega_b}{C_{fi}}i_{lqi}\\
        & -\frac{K_{Pci}K_{Pvi}n_{Qi}}{C_{fi}L_{fi}}Q_i+\frac{K_{Pci}K_{Ivi}}{C_{fi}L_{fi}}\phi_{di}\\
        & +\frac{K_{Ici}}{C_{fi}L_{fi}}\gamma_{di}+\frac{1}{C_{fi}L_{ci}}v_{bdi}
    \end{aligned}\label{eq_lf2}
\end{align}
\begin{align}
    L_{\boldsymbol{g}_i}L_{\boldsymbol{F}_i}h_i(\boldsymbol{x}_i) = \frac{K_{Pci}K_{Pvi}}{C_{fi}L_{fi}}\label{eq_lglf}
\end{align}
where $\omega_b$ is the rated frequency of the MG; $v_{bdi}$ represents $q$-axis voltage at the connection bus between DG and MG; $i_{ldi}, i_{lqi}$ denote the $d$-axis and $q$-axis currents of filter inductance; $K_{Pvi}, K_{Ivi}$ and $K_{Pci}, K_{Ici}$ denote the proportional, integral gains of voltage and current control loops respectively; $R_{fi}, L_{fi}, C_{fi}$ denote resistance, inductance and capacitance values of LC filter; $R_{ci}, L_{ci}$ denote output resistance and inductance values.

From (\ref{eq_dv})-(\ref{eq_lglf}), it is clear that the intermediate variables $\dot{v}_{odi}$, $L_{\boldsymbol{F}_i}^{2}h_i(\boldsymbol{x}_i)$ and $L_{\boldsymbol{g}_i}L_{\boldsymbol{F}_i}h_i(\boldsymbol{x}_i)$ mainly consist of three different parts, measurable variables, immeasurable variables and parameters. To fully investigate the effects of the existing disturbances, we consider three corresponding uncertain factors: measurement noise, exogenous disturbance and parameter perturbation.

For the secondary voltage control of MG, the influence of measurement noise cannot be ignored as such noise can be amplified by inductance and capacitance values that are relatively small (order of magnitude: $10^{-3}$ or $10^{-6}$). Exogenous disturbance from immeasurable variables and parameter perturbation can bring deviation to $L_{\boldsymbol{F}_i}^{2}h_i(\boldsymbol{x}_i)$ and $L_{\boldsymbol{g}_i}L_{\boldsymbol{F}_i}h_i(\boldsymbol{x}_i)$, thus reducing system controlling accuracy. %For those measurable variables such as voltages and currents inside inverters, a large number of measurement units need to be deployed for direct measurements. Moreover, the performance of controller would be affected due to the existing measuring and transmission noise. Parameter perturbation is also inevitable due to the fact that line parameters deviate after long-period use. Immeasurable variables will directly bring deviation, thus reducing system controlling accuracy.\textcolor{red}{(????)}%Traditionally, noise and perturbation may not exert tremendous influence on %data accuracy and 
%%control performance in the systems where control commands are derived from measurements directly. However, in the MG secondary voltage, the effects of measurement noise and parameter perturbation cannot be ignored because such disturbance can be amplified by inductance and capacitance values that are relatively small (order of magnitude: $10^{-3}$ or $10^{-6}$) \textcolor{red}{(will not be amplified in other systems????)}. Together with the influence of immeasurable variables, the system operational robustness of MG under distributed voltage control is not satisfactory.
Moreover, for those measurable variables such as voltages and currents inside the inverters, if the intermediate variables by (\ref{eq_dv})-(\ref{eq_lglf}) are introduced in the control system, a large number of measurement units need to be deployed for direct measurements. %which also leads to large measurement units deployment expense.
Therefore, to overcome negative effects of the aforementioned disturbance, an extended state observer is employed to observe accurate values of $\dot{v}_{odi}$, $L_{\boldsymbol{F}_i}^{2}h_i(\boldsymbol{x}_i)$ and $L_{\boldsymbol{g}_i}L_{\boldsymbol{F}_i}h_i(\boldsymbol{x}_i)$.

The linearized model considering system disturbance for the $i$th DG  (\ref{eq_Linear}) can be extended as:
\begin{align}
    \left\{
    \begin{aligned}
    \dot{y}_{i,1} &= y_{i,2}\\
    \dot{y}_{i,2} &= \xi_i+g_{i,0}u_i\\
    \dot{\xi}_i &= \psi_i\\
    y_i &= y_{i,1}= v_{odi}
    \end{aligned}
    \right.\label{eq_extend_model}
\end{align}
where
\begin{align}
    g_i &= g_{i,0}+\Delta g_i = L_{\boldsymbol{g}_i}L_{\boldsymbol{F}_i}h_i(\boldsymbol{x}_i)\\
    \xi_i &= L_{\boldsymbol{F}_i}^{2}h_i(\boldsymbol{x}_i)+\Delta g_i u_i
\end{align}
$[y_{i,1}\ y_{i,2}]^T$ is the original state vector, while $\xi_i$ is the extend state; $g_{i,0}$ and $\Delta g_i$ denote nominal value and the deviation caused by parameter perturbation of $g_i$ respectively.

Through introducing $\xi_i$, the effect of the uncertainty in $g_i$ is removed by using the constant nominal value. More importantly, the extended state $\xi_i$ represents the sum of DG inner control loops' dynamics and total uncertainties caused by exogenous disturbance, parameter perturbation and the measurement noise. 
%reflecting on the system model. 
As a result, the problem of calculating the three variables (\ref{eq_dv})-(\ref{eq_lglf}) is converted to obtain the values of $\xi_i$ and $y_{i,2}$, which can be directly obtained as the state variables in the system (\ref{eq_extend_model}) by filtering out the influence of zero-mean and high-frequency measurement noise.

\subsection{Multi-Disturbance Resilient Extended State Observer Design Based on Kalman-Bucy Filter} \label{subsection_ESKBF_design}
The extended state system model (\ref{eq_extend_model}) can be written in a matrix form:
\begin{align}
\left\{
    \begin{aligned}
        {\left[ \begin{array}{c}
        \dot{\boldsymbol{x}}_{v,i} \\
        \dot{\xi}_{i}
        \end{array} 
        \right ]}
        = & {\left[\begin{array}{cc}
        \mathbf{A} & \mathbf{E}\\ \mathbf{0} & 0
        \end{array} 
        \right ]}
        {\left[ \begin{array}{c}
        \boldsymbol{x}_{v,i} \\ \xi_{i}
        \end{array} 
        \right ]}
        + {\left[ \begin{array}{c}
        \mathbf{B} \\  0
        \end{array} 
        \right ]}u_i\\
        &+ {\left[ \begin{array}{c}
        \mathbf{0} \\ \psi_{i}
        \end{array} 
        \right ]}
        + \boldsymbol{w}_{i}\\
        \boldsymbol{y}_i =& {\left[ \begin{array}{cc}
        \mathbf{C} & 0
        \end{array} 
        \right ]}
        {\left[ \begin{array}{c}
        \boldsymbol{x}_{v,i} \\ {\xi}_{i}
        \end{array} 
        \right ]}
        + \boldsymbol{\nu}_i
    \end{aligned}
    \right.\label{eq_ex_matrix}
\end{align}
    where $\boldsymbol{x}_{v,i}=[v_{odi}\ \dot{v}_{odi}]^T$ denotes original state vector in the second-order system; $\boldsymbol{w}_i=[\boldsymbol{w_x}_{,i}\ 0]^T$ and $\boldsymbol{\nu}_i$ respectively represent the process noise and the observation noise; the corresponding constant matrices are 
\begin{align*}
    \mathbf{A}={\left[\begin{array}{cc}
    0 & 1   \\  0 & 0
    \end{array} 
    \right ]},
    \mathbf{B}&={\left[\begin{array}{cc}
    0   \\ g_{i,0}
    \end{array} 
    \right ]},
    \mathbf{C}={\left[\begin{array}{cc}
    1 & 0
    \end{array}
    \right ]},
    \mathbf{E}&={\left[\begin{array}{cc}
    0   \\ 1
    \end{array}
    \right ]}.
\end{align*}

To simplify the observer design, system (\ref{eq_ex_matrix}) can be expressed as
\begin{align}
\left\{
\begin{aligned}
    \dot{\boldsymbol{x}}_{{\rm ex},i} =& \mathbf{A}_{\rm ex}{\boldsymbol{x}}_{{\rm ex},i}+\mathbf{B}_{\rm ex}u_i+\boldsymbol{\Psi}_i+\boldsymbol{w}_i\\
    \boldsymbol{y}_i =& \mathbf{C}_{\rm ex}{\boldsymbol{x}}_{{\rm ex},i}+\boldsymbol{\nu}_i
    \end{aligned}
    \right.\label{eq_observer_model}
\end{align}
where
\begin{align*}
    {\boldsymbol{x}}_{{\rm ex},i} = {\left[ \begin{array}{c}
    \boldsymbol{x}_{v,i} \\ \xi_{i}
    \end{array} 
    \right ]},
    \boldsymbol{\Psi}_i = {\left[ \begin{array}{c}
    \mathbf{0} \\ \psi_{i}
    \end{array} 
    \right ]},
\end{align*}
\begin{align*}
    \mathbf{A}_{\rm ex} = {\left[\begin{array}{cc}
    \mathbf{A} & \mathbf{E}\\ \mathbf{0} & 0
    \end{array} 
    \right ]},
    \mathbf{B}_{\rm ex} = {\left[ \begin{array}{c}
    \mathbf{B} \\  0
    \end{array} 
    \right ]},
    \mathbf{C}_{\rm ex} = {\left[ \begin{array}{cc}
    \mathbf{C} & 0
    \end{array} 
    \right ]}.
\end{align*}

{\emph{Assumption 3.1:}} The process noise and the observation noise are energy limited:
\begin{align}
    & E\{\boldsymbol{w_x}_{,i}\boldsymbol{w_x}_{,i}^T\}\leq \boldsymbol{Q_x}\\
    & E\{\boldsymbol{\nu}_{i}\boldsymbol{\nu}_{i}^T\}\leq \boldsymbol{R_x}
\end{align}
where $\boldsymbol{Q_x}>0$ and $\boldsymbol{R_x}>0$; $\boldsymbol{Q_x}$ and $\boldsymbol{R_x}$ are upper bounded.

{\emph{Assumption 3.2:}} The unknown dynamics are upper bounded: $E\{\psi_{i}\psi_{i}^T\}\leq Q_{\xi}$ and $Q_{\xi}>0$.

To design multi-disturbance resilient observer based on the model~(\ref{eq_observer_model}), a novel observer, combining extended stated observer with Kalman-Bucy filter, can be fomulated as follows:
\begin{align}
    \left\{
    \begin{aligned}
        \hat{\dot{\boldsymbol{x}}}_{{\rm ex},i} =& \mathbf{A}_{\rm ex}{\hat{\boldsymbol{x}}}_{{\rm ex},i}+\mathbf{B}_{\rm ex}u_i+\boldsymbol{K}_i(\boldsymbol{y}_i-{\mathbf{C}_{\rm ex}}{\boldsymbol{x}}_{{\rm ex},i})\\
        \dot{\boldsymbol{P}_i} =& \mathbf{A}_{\rm ex}{\boldsymbol{P}_i}+{\boldsymbol{P}_i}{\mathbf{A}_{\rm ex}^T}-{\boldsymbol{K}_i}{\mathbf{C}_{\rm ex}}{\boldsymbol{P}_i}+{\boldsymbol{Q}_i}\\
        {\boldsymbol{K}_i} =& {\boldsymbol{P}_i}{\mathbf{C}_{\rm ex}^T}{\boldsymbol{R}_i^{-1}}
    \end{aligned}
    \right.
\end{align}
where
\begin{align*}
    {\boldsymbol{Q}_i} = {\left[\begin{array}{cc}
    {\boldsymbol{Q_x}} & \mathbf{0}\\ \mathbf{0} & Q_{\xi}
    \end{array} 
    \right ]},
    {\boldsymbol{R}_i} = {\boldsymbol{R_x}}
\end{align*}
It is noted that the estimation error of the initial state is bounded, and the initial parameter $\boldsymbol{P}_{i,0}$ is selected by satisfying $E\{({\boldsymbol{x}_{{\rm ex},i}-\hat{\boldsymbol{x}}}_{{\rm ex},i})({\boldsymbol{x}_{{\rm ex},i}-\hat{\boldsymbol{x}}}_{{\rm ex},i})^T\} \leq \boldsymbol{P}_{i,0}$, %which is quite general by setting a sufficient large $\boldsymbol{P}_{i,0}$.
Furthermore, the conditions in \emph{Assumption 3.1} and \emph{Assumption 3.2} can be met due to the limited power of the systems in practice~\cite{he_distributed_2018}. Thus, parameters $\boldsymbol{Q}_i, \boldsymbol{R}_i$ can be selected accordingly, supported by  the simple dynamics of the extended system (\ref{eq_ex_matrix}).

\section{Distributed Robust Fast Terminal Sliding Mode Secondary Voltage Control}\label{section_vol_ctrl}
In this section, a distributed fast terminal sliding mode (FTSM) control strategy is introduced to select the appropriate voltage control input $V_{ni}$ to guarantee the voltage magnitude of DGs following the voltage reference $v_{ref}$, $v_{o,magi}=v_{odi} \to v_{ref}$. The proposed FTSM control strategy can address the consensus tracking problem of linear second-order system (\ref{eq_sec_ord}) with the reference $v_{ref}$. Then, to realize the trade-off between voltage regulation and accurate reactive power sharing, the extension of the proposed FTSM control strategy is also provided.

\subsection{Voltage Regulation}
The corresponding reference with regard to the second-order system (\ref{eq_sec_ord}) is 
\begin{align}
    \left\{
    \begin{aligned}
    y_0 =& v_{ref}\\
    \dot{y}_0 =& 0
    \end{aligned}
    \right.
\end{align}
which is commonly accessible to one DG node, thus the tracking errors of local neighborhood state for the $i$th DG node can be denoted as
\begin{align}
    \left\{
    \begin{aligned}
        e_{i,1} =& \sum_{j \in N_{i}}{a_{ij}(y_{i,1}-y_{j,1})}+b_{i}(y_{i,1}-y_{0})\\
        e_{i,2} =& \sum_{j \in N_{i}}{a_{ij}(y_{i,2}-y_{j,2})}+b_{i}y_{i,2}
    \end{aligned}
    \right.\label{eq_tracking_error}
\end{align}

To solve the above tracking control problem,
%where the errors are described as (\ref{eq_tracking_error}), 
the following FTSM surface is designed
\begin{align}
    s_i = e_{i,2}+ce_{i,1}^{m/n}+de_{i,1}^{p/q}\label{eq_sliding_surface}
\end{align}
where $c,d>0$, and $m,n,p,q$ are positive odd integers satisfying $m>n,p<q$. The nonlinear terminal attractors $e_{i,1}^{m/n}$ and  $e_{i,1}^{p/q}$ are applied to improve the convergence rate. More especially, the term $e_{i,1}^{p/q}$ can improve the convergence rate when the system is close to the equilibrium~\cite{ni_fast_2017}.

To solve the sliding mode surface~(\ref{eq_sliding_surface}), nonlinear function ${\rm sig}(x)^{a}={\rm sgn}(x)\left|x\right|^{a}$ with signum ${\rm sgn}(x)$~\cite{zuo_non-singular_2015} is employed to design the control law for the system (\ref{eq_sec_ord}) as follows:
\begin{align}
\begin{aligned}
    z_{i} =& \left(\sum_{j \in N_{i}}{a_{ij}+b_{i}}\right)^{-1}\Bigg[\sum_{j \in N_{i}}{a_{ij}z_{j}}-\alpha{\rm sig}(s_{i})^2\\
    & -\beta{\rm sgn}(s_{i})-\left(c\frac{m}{n}e_{i,1}^{m/n-1}+d\frac{p}{q}e_{i,1}^{p/q-1}\right)e_{i,2}\Bigg]
    \end{aligned}\label{eq_control_law}
\end{align}
where $\alpha,\beta>0$.

Using the control law \eqref{eq_control_law} under the sliding model surface \eqref{eq_sliding_surface}, the distributed secondary voltage regulation problem can be solved with the stability guaranteed. 

\begin{IEEEproof}
To verify the system stability under the control law (\ref{eq_control_law}), consider the following Lyapunov candidate function:
\begin{align}
    V = \frac{1}{2}\sum_{i=1}^{N}{s_{i}^{2}}\label{eq_Lyapunov_function}
\end{align}
and the time derivative of $V$ can be obtained as
\begin{align}
\begin{aligned}
    \dot{V} & = \sum_{i=1}^{N}{s_{i}\dot{s}_{i}}\\
    & = \sum_{i=1}^{N}{s_{i}\Bigg[\dot{e}_{i,2}+\left(c\frac{m}{n}e_{i,1}^{m/n-1}+d\frac{p}{q}e_{i,1}^{p/q-1}\right)e_{i,2}}\Bigg]\\
    & = \sum_{i=1}^{N}{s_{i}\left(-\alpha{\rm sig}(s_{i})^2-\beta{\rm sgn}(s_{i})\right)}\\
    & = -\alpha\sum_{i=1}^{N}{\left(s_{i}^{2}\right)^{\frac{3}{2}}}-\beta\sum_{i=1}^{N}{\left(s_{i}^{2}\right)^{\frac{1}{2}}}
\end{aligned}
\end{align}
Based on \emph{Lemma 3.3} and \emph{Lemma 3.4} in~\cite{zuo_distributed_2016_1}, we obtain
\begin{align}
\begin{aligned}
    \dot{V} & \leq -\alpha N^{-\frac{1}{2}}\left(\sum_{i=1}^{N}{s_{i}^{2}}\right)^{\frac{3}{2}}-\beta\left(\sum_{i=1}^{N}{s_{i}^{2}}\right)^{\frac{1}{2}}\\
    & \leq -\alpha N^{-\frac{1}{2}}\left(2V\right)^{\frac{3}{2}}-\beta\left(2V\right)^{\frac{1}{2}}
\end{aligned}
\end{align}
Since $V>0,\dot{V}<0$, following  \emph{Lemma 4.1} in~\cite{zuo_distributed_2016_1}, the convergence of (\ref{eq_Lyapunov_function})  towards 0 is guaranteed. Thus, $s=0$ will be maintained under the control law (\ref{eq_control_law}). %and the proof is completed.

After the sliding mode surface has been reached, the dynamics of tracking errors $e_{i,1}$ can be described by
\begin{align}
    \begin{aligned}
        &s_i = \dot{e}_{i,1}+ce_{i,1}^{m/n}+de_{i,1}^{p/q} = 0\\
        &\Rightarrow{\dot{e}_{i,1}=-ce_{i,1}^{m/n}-de_{i,1}^{p/q}}
    \end{aligned}\label{eq_FTSM_surface}
\end{align}
Based on the stability proof of system (\ref{eq_FTSM_surface}) in \cite{ni_fast_2017}, the stability of the designed FTSM secondary voltage control can been proved.
\end{IEEEproof}

\subsection{Trade-off between Voltage Regulation and Reactive Power Sharing}\label{subsection_trade-off}

The exact voltage restoration and accurate reactive power sharing cannot be achieved simultaneously due to the line impedance effect~\cite{dehkordi_distributed_2017}, except for a perfectly symmetric configuration~\cite{zuo_distributed_2016}. However, the exact voltage regulation and accurate reactive power sharing could compromise with each other based on the practical circumstances.
For the cases that sensitive loads require operation at the nominal voltage or the overloading of DGs is not the primary concern, voltage regulation should be prioritised \cite{bidram_distributed_2013,zuo_distributed_2016,dehkordi_distributed_2017,pilloni_robust_2018,ge_event-triggered_2020}. However, if the concerned system has low ratings of DGs, small electrical distances between DGs or limited capacitive compensation, the reactive power sharing needs to be maintained to prevent overloading \cite{gu_nonlinear_2017,zhang_distributed_2017,wang_cyber-physical_2019,simpson-porco_secondary_2015}. Thus, in this subsection, the extension of the proposed secondary control scheme is provided for the case that a trade-off between voltage regulation and accurate reactive power sharing needs to be achieved.

The FTSM surface \eqref{eq_sliding_surface} and the control law \eqref{eq_control_law} can be respectively modified as
\begin{align}
    &\begin{aligned}s_i = &e_{i,2}+ce_{i,1}^{m/n}+de_{i,1}^{p/q}+c_{q}e_{qi}^{m1/n1}+d_{q}e_{qi}^{p1/q1}\\
    &e_{qi}=\sum_{j \in N_{i}}{a_{ij}(n_{Qi}Q_{i}-n_{Qj}Q_{j})}
    \end{aligned}\label{eq_sliding_surface2}\\
    &\begin{aligned}
    z_{i} =& \left(\sum_{j \in N_{i}}{a_{ij}+b_{i}}\right)^{-1}\Bigg[\sum_{j \in N_{i}}{a_{ij}z_{j}}-\alpha{\rm sig}(s_{i})^2\\
    & -\beta{\rm sgn}(s_{i})-\left(c\frac{m}{n}e_{i,1}^{m/n-1}+d\frac{p}{q}e_{i,1}^{p/q-1}\right)e_{i,2}\Bigg]
    \end{aligned}\label{eq_control_law2}
\end{align}
where $c_q,d_q>0$, and $m1,n1,p1,q1$ are positive odd integers satisfying $m1>n1,p1<q1$ similarly. 

To express the control trade-off, the Laplacian matrix of the distributed system is redefined as $\overline{\mathcal{L}}_{V}$ and $\mathcal{L}_{Q}$:
\begin{align}
    &\left\{\begin{aligned}
        &\overline{\mathcal{L}}_{V}=\mathcal{L}_{V}+\mathcal{B}_{V}\\
        &\left[e_{i,1}\right]_{N\times1}=\mathcal{L}_{V}\left[y_{i,1}\right]_{N\times1}+\mathcal{B}_{V}\left[y_{i,1}-y_{0}\right]_{N\times1}\\
        &\left[e_{i,2}\right]_{N\times1}=\mathcal{L}_{V}\left[y_{i,2}\right]_{N\times1}+\mathcal{B}_{V}\left[y_{i,2}\right]_{N\times1}
    \end{aligned}\right.\\
    &\left[e_{qi}\right]_{N\times1}=\mathcal{L}_{Q}\left[n_{Qi}Q_{i}\right]_{N\times1}
\end{align}
where $\left[*\right]_{N\times1}$ denotes the column vector composed of states of all DG units. If $\overline{\mathcal{L}}_{V}\neq\textbf{0},\mathcal{L}_{Q}=\textbf{0}$, the control system is the same as that only emphasizes voltage regulation. If $\overline{\mathcal{L}}_{V}=\textbf{0},\mathcal{L}_{Q}\neq\textbf{0}$, the control system is the same as that only emphasizes accurate reactive power sharing. However, $\overline{\mathcal{L}}_{V}=\textbf{0}$ could lead to poor voltage regulation. Let $\mathcal{L}_{V}=0,\mathcal{L}_{Q}\neq\textbf{0},\mathcal{B}_{V}\neq\textbf{0}$, the accurate reactive power sharing is guaranteed, while the voltages are regulated all around the reference. Regarding this trade-off, how to select an optimal $\mathcal{B}_{V}$ would be an interesting problem, and we will consider this by an optimization algorithm in the future work.

\section{Controller Implementation for MGs}\label{section_implement}
\begin{figure}[!t]
\centering
\includegraphics[width=0.45\textwidth]{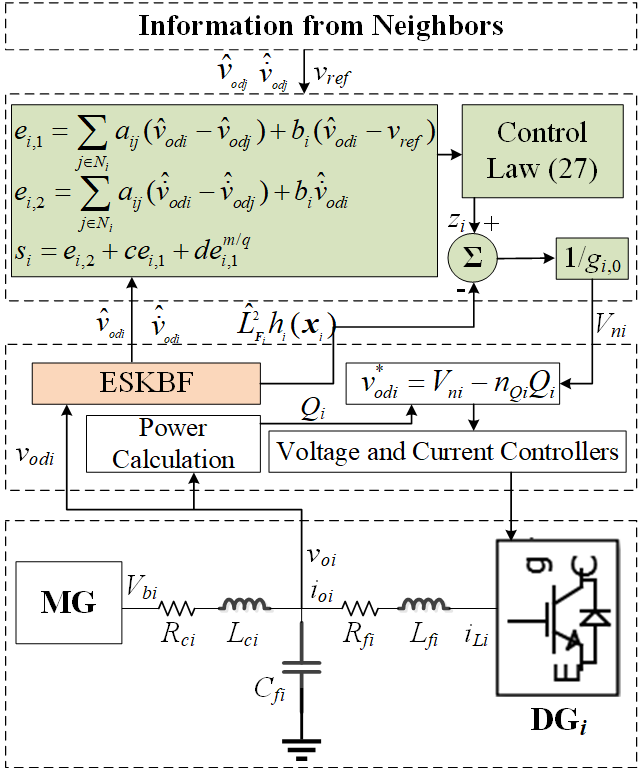}
\caption{Block diagram of the proposed distributed robust FTSM voltage control.}
\label{fig_implementation}
\end{figure}

The diagram of the proposed ESKBF-based distributed robust voltage control of MGs is detailed in Fig.~\ref{fig_implementation}. The ESKBF only requires the local voltage information to observe the corresponding information related to the inverter's dynamics. Once the observed information against disturbance and uncertainties is obtained and transmitted to its neighbors, the nominal control input $V_{ni}$ can be updated to respond to system operation changes through the FTSM-based secondary voltage control law accordingly. Similarly, if the reactive power sharing is considered, the control implementation should also be modified as that analyzed in Section \ref{subsection_trade-off}.

\section{Simulation and Experimental Results}\label{section_simulation}

\begin{figure}[!b]
\centering
\includegraphics[width=2.5in]{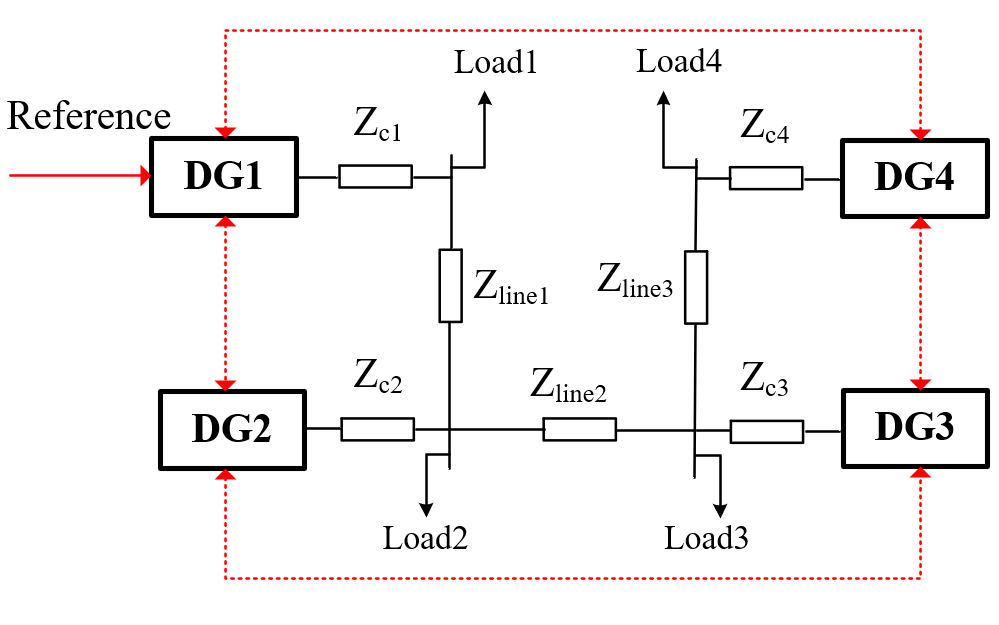}
\caption{Diagram of a 4-bus MG.}
\label{fig_simulation_diagram}
\end{figure}

In this section, to verify the effectiveness of the proposed ESKBF-based distributed FTSM secondary voltage control method, both simulation and experimental studies are developed.
More specifically, the proposed control scheme is firstly tested on a 4-DG islanded MG. Then, the scalability and practical performance of the proposed method are evaluated by a modified IEEE 37 bus system and an experimental MG testbed respectively. 

\begin{table}[!t]
\renewcommand{\arraystretch}{1.3}
\caption{Parameters of the Tested 4-bus MG System}
\label{table_parameter}
\centering
    \begin{tabular}{ccccc}
        \hline\hline
        \multicolumn{2}{c}{}               & DG1                        & DG2                   & DG3 \& DG4 \\ \hline
        \multicolumn{2}{c}{DG power ratings}     &40kW, 30kVar        &27kW, 20kVar    &20kW, 15kVar \\
        \multirow{11}{*}{DGs}   &$m_P$      &$6.28\times10^{-5}$        &$9.42\times10^{-5}$    &$12.56\times10^{-5}$ \\
                                &$n_{Q}$    &$0.5\times10^{-3}$         &$0.75\times10^{-3}$    &$1\times10^{-3}$ \\
                                &$R_{f}$    &0.1 $\Omega$               &0.1 $\Omega$           &0.1 $\Omega$ \\
                                &$L_{f}$    &1.35 mH                    &1.35 mH                &1.35 mH \\
                                &$C_{f}$    &47$ \mu$F                  &47$ \mu$F              &47 $\mu$F \\
                                &$R_{c}$    &0.02 $\Omega$              &0.02 $\Omega$          &0.02 $\Omega$ \\
                                &$L_{c}$    &2 mH                       &2 mH                   &2 mH \\
                                &$K_{Pv}$   &0.05                       &0.05                   &0.1 \\
                                &$K_{Iv}$   &390                        &390                    &420 \\
                                &$K_{Pc}$   &10.5                       &10.5                   &15 \\
                                &$K_{Ic}$   &$1.6\times10^{4}$          &$1.6\times10^{4}$      &$2\times10^{4}$ \\ \hline
        \multirow{3}{*}{Lines}  &Line1   &\multicolumn{3}{c}{${\rm R=0.23~\Omega,~L=318~\mu H}$} \\
                                &Line2   &\multicolumn{3}{c}{${\rm R=0.35~\Omega,~L=1847~\mu H}$} \\
                                &Line3   &\multicolumn{3}{c}{${\rm R=0.23~\Omega,~L=318~\mu H}$} \\ \hline
        \multirow{4}{*}{RL Loads} &Load1 &\multicolumn{3}{c}{${\rm R=4~\Omega,~L=9.6~mH}$} \\                                                                             
                                &Load2   &\multicolumn{3}{c}{${\rm R=8~\Omega,~L=12.8~mH}$} \\
                                &Load3   &\multicolumn{3}{c}{${\rm R=6~\Omega,~L=12.8~mH}$} \\
                                &Load4   &\multicolumn{3}{c}{${\rm R=12~\Omega,~L=25.6~mH}$} \\ \hline
        \multicolumn{2}{c}{\multirow{3}{*}{Control Parameters}}  &\multicolumn{3}{c}{$c=600, m=13, n=11$} \\
        \multicolumn{2}{c}{}             & \multicolumn{3}{c}{$d=100, p=3, q=5$} \\
        \multicolumn{2}{c}{}             & \multicolumn{3}{c}{$\alpha=100, \beta=400, v_{ref}=311~{\rm V}$} \\ \hline\hline
    \end{tabular}
\end{table}

\begin{figure}[!b]
\centering
\includegraphics[width=0.45\textwidth]{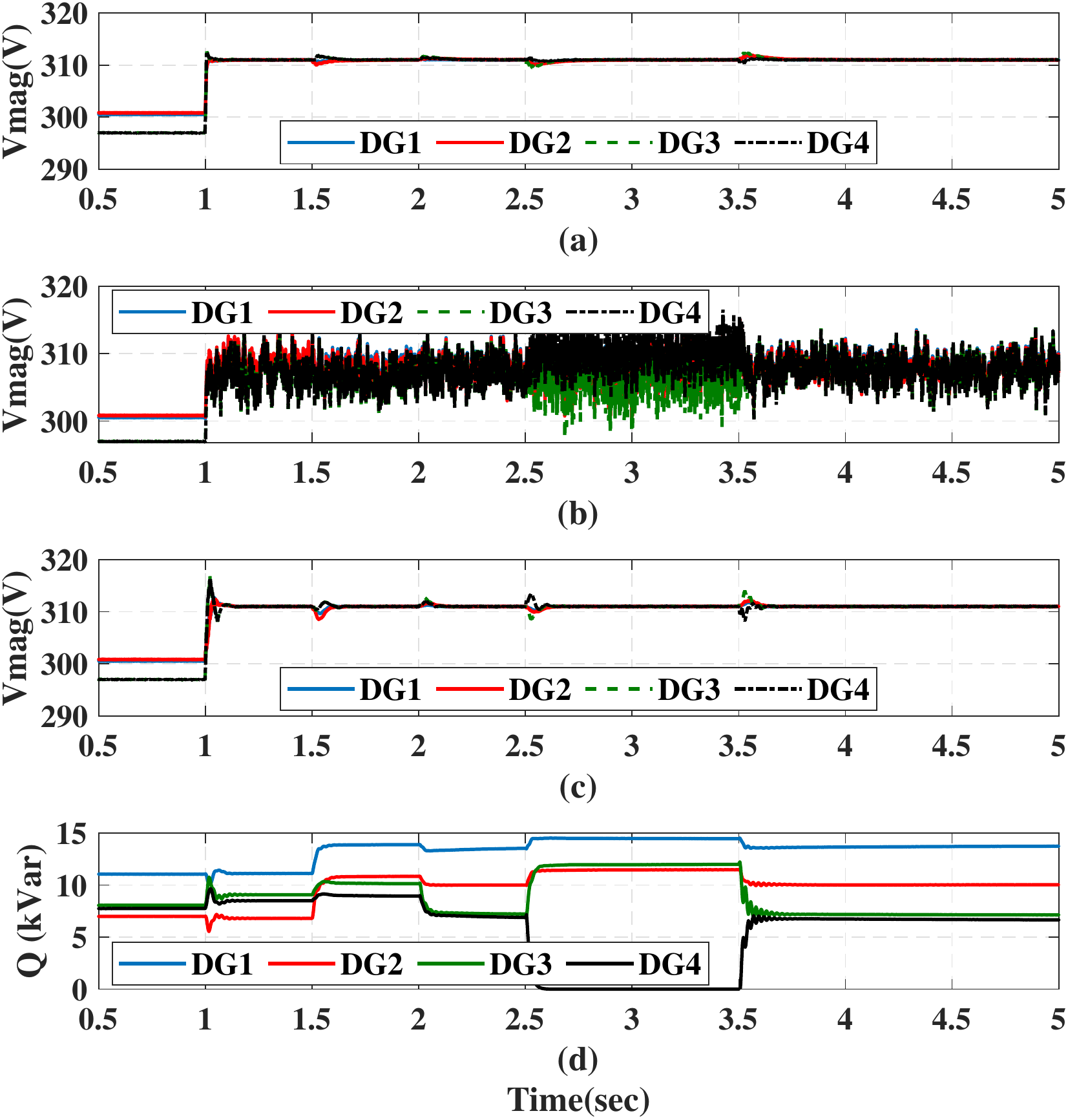}
\caption{General performance evaluation of ESKBF-based distributed voltage control: (a) noise-free environment without observer, (b) noise-containing environment without observer, (c) noise-containing environment with ESKBF, (d) reactive power output in the noise-containing environment with ESKBF.}\label{fig_general_comparison}
\end{figure}

\begin{figure*}[!t]
\centering
\includegraphics[width=0.85\textwidth]{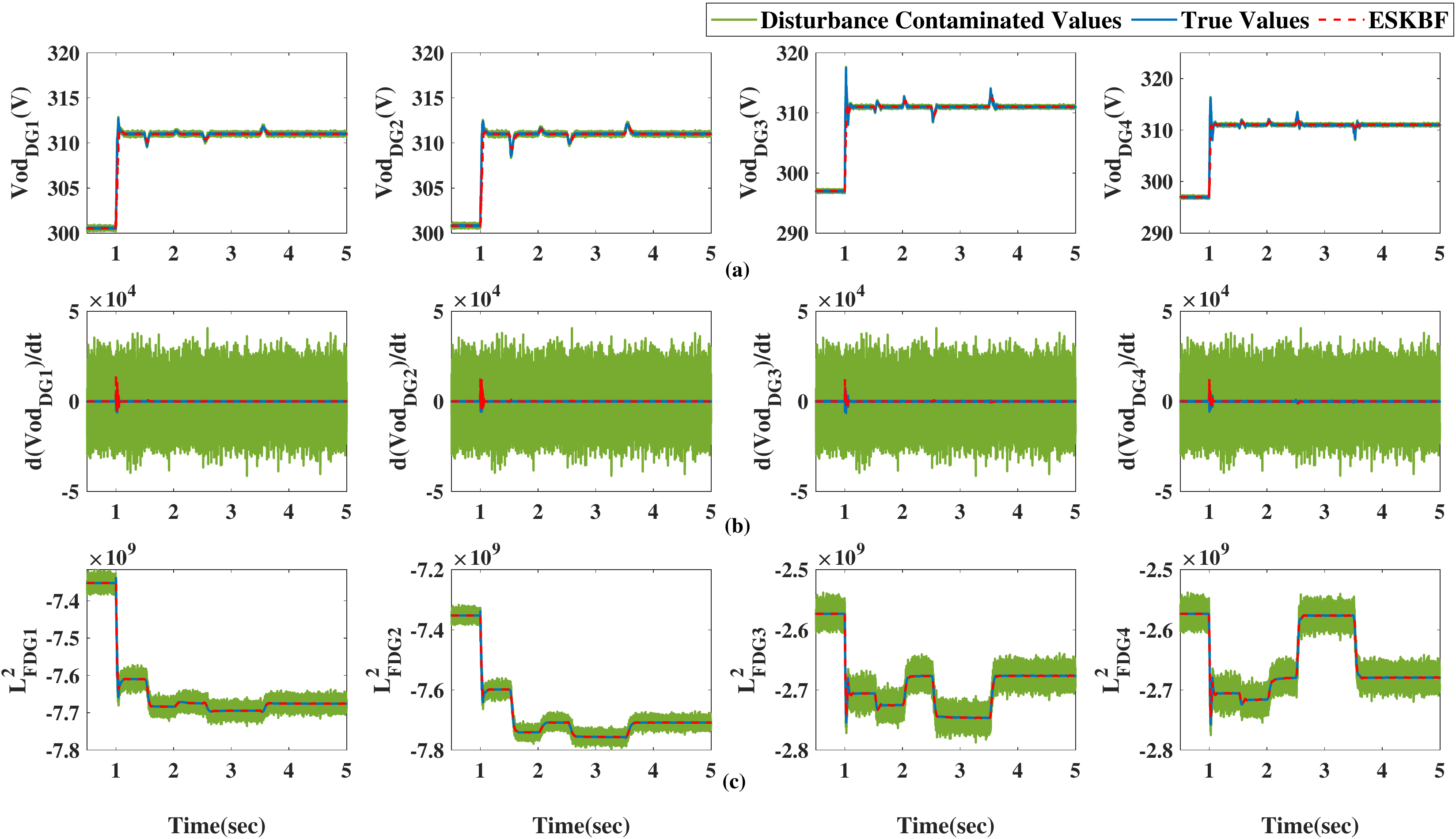}
\caption{ESKBF-based observer performance evaluation: (a) $v_{odi}$, (b) $\dot{v}_{odi}$, (c) $L_{\boldsymbol{F}_i}^{2}h_i(\boldsymbol{x}_i)$.}\label{fig_General_ESKBF}
\end{figure*}

The 4-DG islanded MG, as shown in Fig.~\ref{fig_simulation_diagram}, is developed in the MATLAB/Simulink and parameters are detailed in Table~\ref{table_parameter}. In this 4-bus MG system, the following simulation scenario is designed to evaluate the performance of the proposed voltage control strategy:

\begin{enumerate}
    \item $t=0.0$ s: simulation initialization period, when only the primary controller is applied with constant control input $V_{ni}=311{\rm V}$, and Load2 is not connected into the MG.
    \item $t=1.0$ s: the proposed ESKBF-based distributed robust FTSM secondary voltage control is activated;
    \item $t=1.5$ s: Load2 is connected into the MG (100\% increment of the load);
    \item $t=2.0$ s: Load3 is decreased (50\% decrement of the load);
    \item $t=2.5$ s: DG4 is disconnected (plugged out);
    \item $t=3.5$ s: DG4 is re-connected (plugged in).
\end{enumerate}

\subsection{General Performance Analysis}\label{subsection_result_general}

To demonstrate the negative effects of ignoring measurement noise when designing the controller, %as analyzed in the subsection~\ref{subsec_control-oriented model}, 
we simulate the MG system operation with the controller designed without considering the measurement noise in both noise-free and noise-containing environments. In the noise-containing environment, the additive measurement noise with $\sigma^{2}=0.01$ 
%and the line parameters with 15\% increment from the values in Table~\ref{table_parameter}~(parameter perturbation) are 
is added throughout the simulation. In the case without noise, the controller accurately regulates the system voltage according to the reference ~(Fig.~\ref{fig_general_comparison}(a)). However, as demonstrated in Fig.~\ref{fig_general_comparison}(b), although the noise amplitude is very small, the voltage control performance is significantly degraded, which is mainly driven by amplification effect of the linearized model.

In the same noise-containing environment, simulation results of the proposed ESKBF-based distributed robust FTSM voltage control are detailed in Fig.~\ref{fig_general_comparison}(c), and the corresponding reactive power output is shown in Fig.~\ref{fig_general_comparison}(d). At the initial phase, the secondary control is not activated, causing a static voltage deviation. Once the proposed secondary control is applied at $t=1$ s, the voltage is restored to its reference. Then, at $t=1.5$ s and $t=2$ s, the control system  rapidly responds to the change of load, and the voltage is accurately regulated to the reference. When DG4 is plugged out and in at $t=2.5$ s and $t=3.5$ s respectively, the voltage restoration can be guaranteed as well. Although some transient still occurs when DG4 is re-connected, the voltage can be rapidly regulated to the reference.

By employing the proposed ESKBF, the negative effects of the disturbance can be eliminated, as shown in Fig.~\ref{fig_General_ESKBF}. More specifically, the impact of ESKBF-based observer is emphasized by the comparisons among true values, ESKBF observed values and disturbance contaminated values that are obtained from indirect measurement. If the MG voltage controller operates without ESKBF, the control performance will degrade as Fig.~\ref{fig_general_comparison}(b), where the voltage fluctuation is undesired and unacceptable.

%{\color{blue} To further demonstrate the effectiveness of proposed ESKBF, we substitute ESKBF with the existing Luenberger-like extended state observer in the control system. Compared to the convenient observation parameters selection of the proposed ESKBF, which has been discussed in subsection~\ref{subsection_ESKBF_design}, empirical parameter selection in~\cite{ge_extended-state-observer-based_2020} will dominate the observation performance because the Luenberger-like extended state observer substantially belongs to the high-gain observer~\cite{madonski_survey_2015}. Without fine tuning of the parameters the observation performance of Luenberger-like extended state observer may become unsatisfactory and the corresponding voltage control performance will degrade accordingly, as shown in Fig.~\ref{fig_conventional_ESO}. %By using inappropriate parameters that probably lead to undesired observation tracking capability, the fluctuation and deviation of system voltage in the MG would occur. 

%\begin{figure}[!t]
%\centering
%\includegraphics[width=0.48\textwidth]{conventional_ESO.eps}
%\caption{Voltage control performance by using the conventional Luenberger-like extended state observer with inappropriate parameters: (a) low observer gain, (b) high observer gain.}\label{fig_conventional_ESO}
%\end{figure}
%}

\subsection{Robustness against Different Disturbance Scenarios}
\begin{figure}[!t]
\centering
\includegraphics[width=0.45\textwidth]{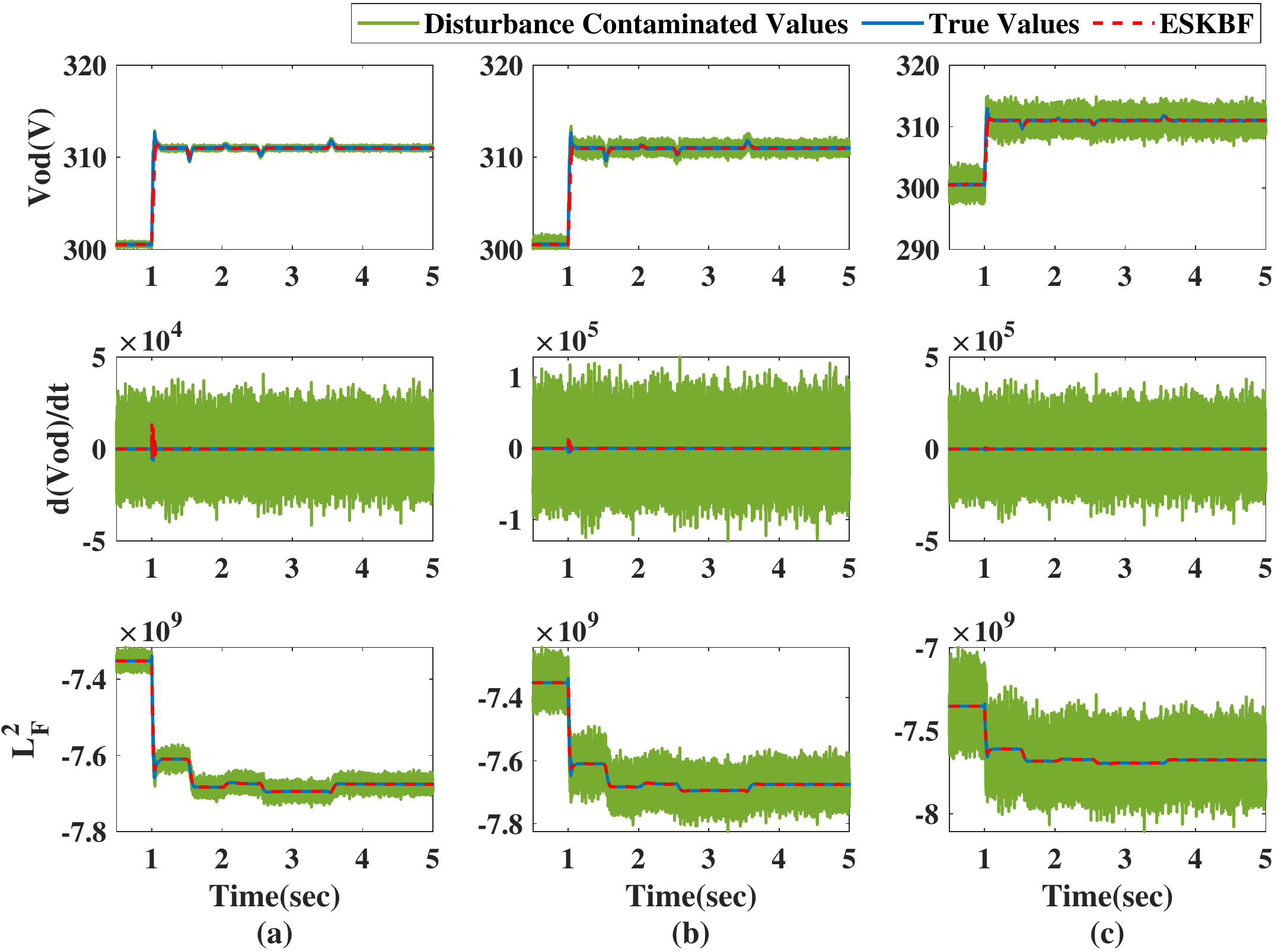}
\caption{Robustness evaluation of ESKBF-based observer: (a) noise $\sigma^{2}=0.01$, (b) noise $\sigma^{2}=0.1$, (c) noise $\sigma^{2}=1$.}\label{fig_Robust_ESKBF}
\end{figure}

\begin{figure}[!t]
\centering
\includegraphics[width=0.45\textwidth]{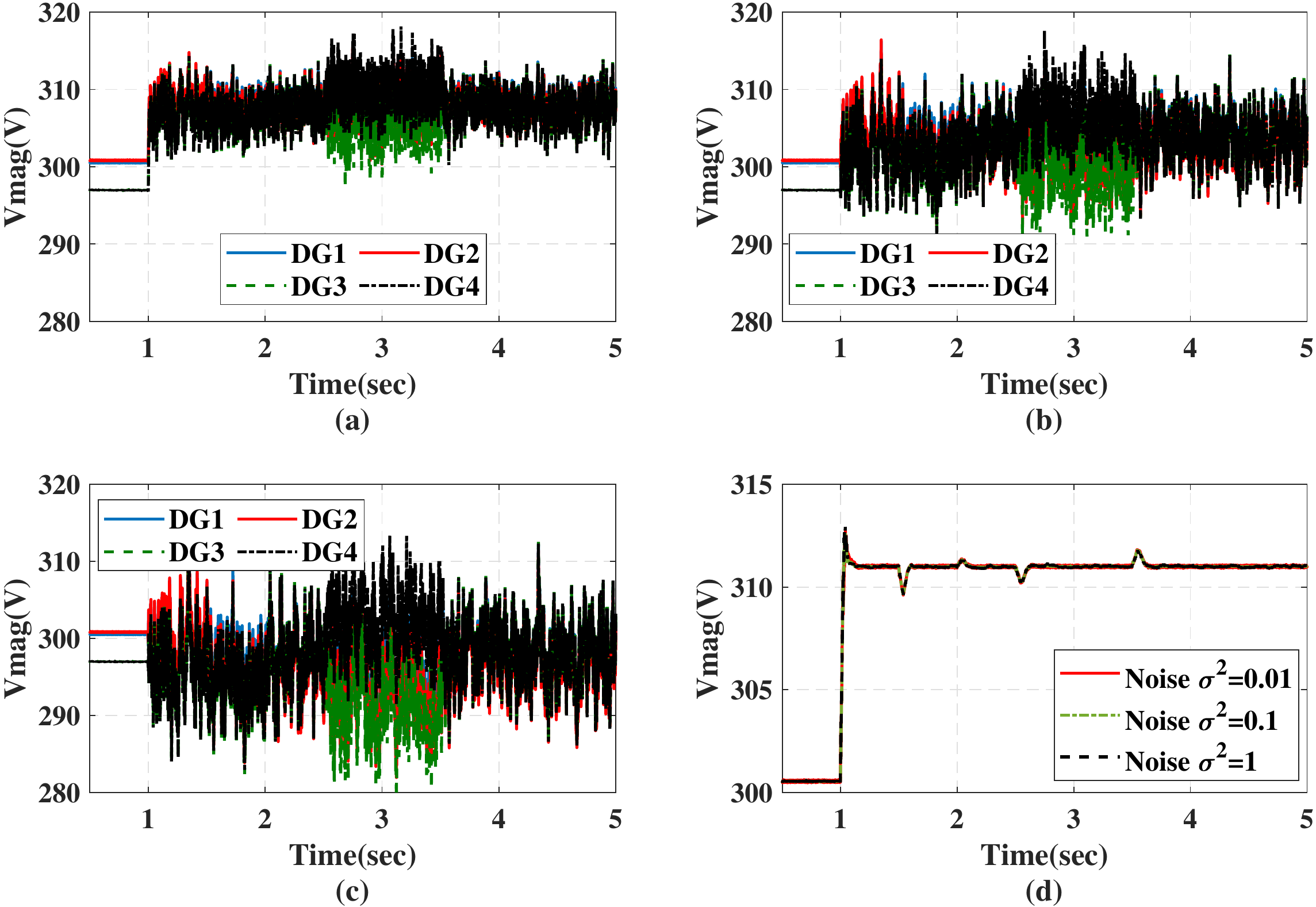}
\caption{Voltage control performance analysis: (a) noise $\sigma^{2}=0.01$, (b) noise $\sigma^{2}=0.1$, (c) noise $\sigma^{2}=1$, (d) voltage comparison of all scenarios.}\label{fig_Robust_voltage}
\end{figure}

To illustrate the robustness of the proposed control method against uncertainties, %based on the scenario where the line parameters are with 15\% increment from the values in Table~\ref{table_parameter},
different levels of measurement noise, $\sigma^{2}=0.01, \sigma^{2}=0.1, \sigma^{2}=1$ are employed in the system. For the sake of simplification, only ESKBF-based observation performance of DG1 is selected, and the corresponding comparisons are shown in Fig.~\ref{fig_Robust_ESKBF}. Although the noise variance is varying, the ESKBF remains effective to filter out the additive noise. Moreover, as shown in Fig.~\ref{fig_Robust_voltage}(d), the proposed ESKBF-based observer enables the voltages being accurately regulated in all cases, demonstrating the robustness of proposed control strategies against unknown level of bounded noise. If ESKBF-based observer is not activated in the secondary voltage controller, the voltages will degrade as Fig.~\ref{fig_Robust_voltage}(a),(b),(c) respectively.

\subsection{Control Performance of Distributed FTSM Secondary Voltage Control }
\begin{figure}[!t]
\centering
\includegraphics[width=0.45\textwidth]{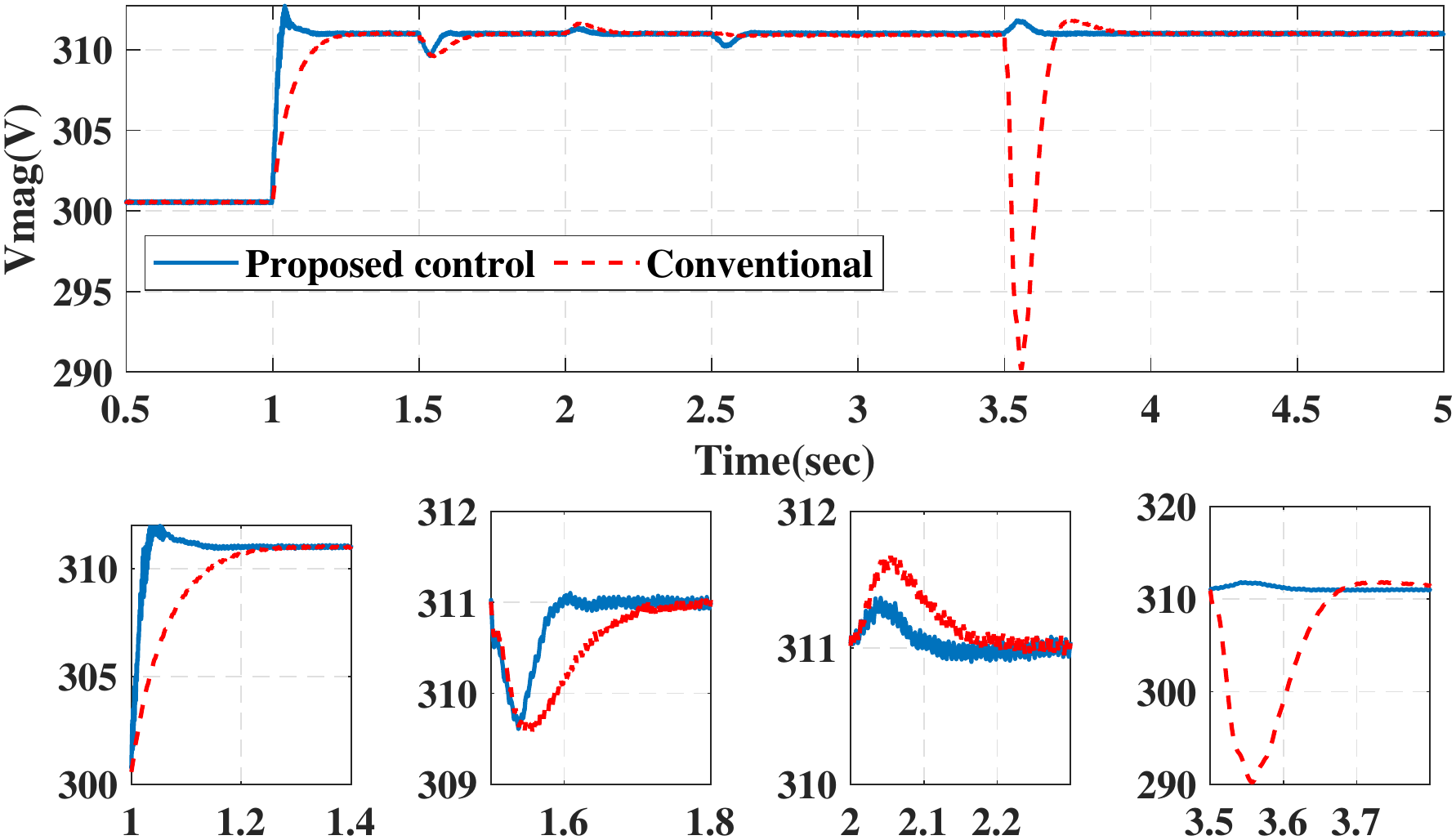}
\caption{Comparison with conventional finite-time control.}\label{fig_Comparison_previous}
\end{figure}

To show the faster convergence of the proposed distributed FTSM secondary voltage control, we compared it with the existing MG distributed finite-time control law~\cite{dehkordi_distributed_2017}. Take DG1's control performance as an example, the comparison is detailed in Fig.~\ref{fig_Comparison_previous}, where both consensus convergence rate and undesired control dynamics are improved by employing the proposed secondary voltage control method, although there may be a slight overshoot when system operation conditions change. It is also worth noting that during plug-in operation at 3.5s, the control performance achieves the most significant improvement, demonstrating the merit of the proposed control framework in supporting plug-and-play operation.

\subsection{Trade-off between Voltage Regulation and Reactive Power Sharing}
\begin{figure}[!t]
\centering
\includegraphics[width=0.48\textwidth]{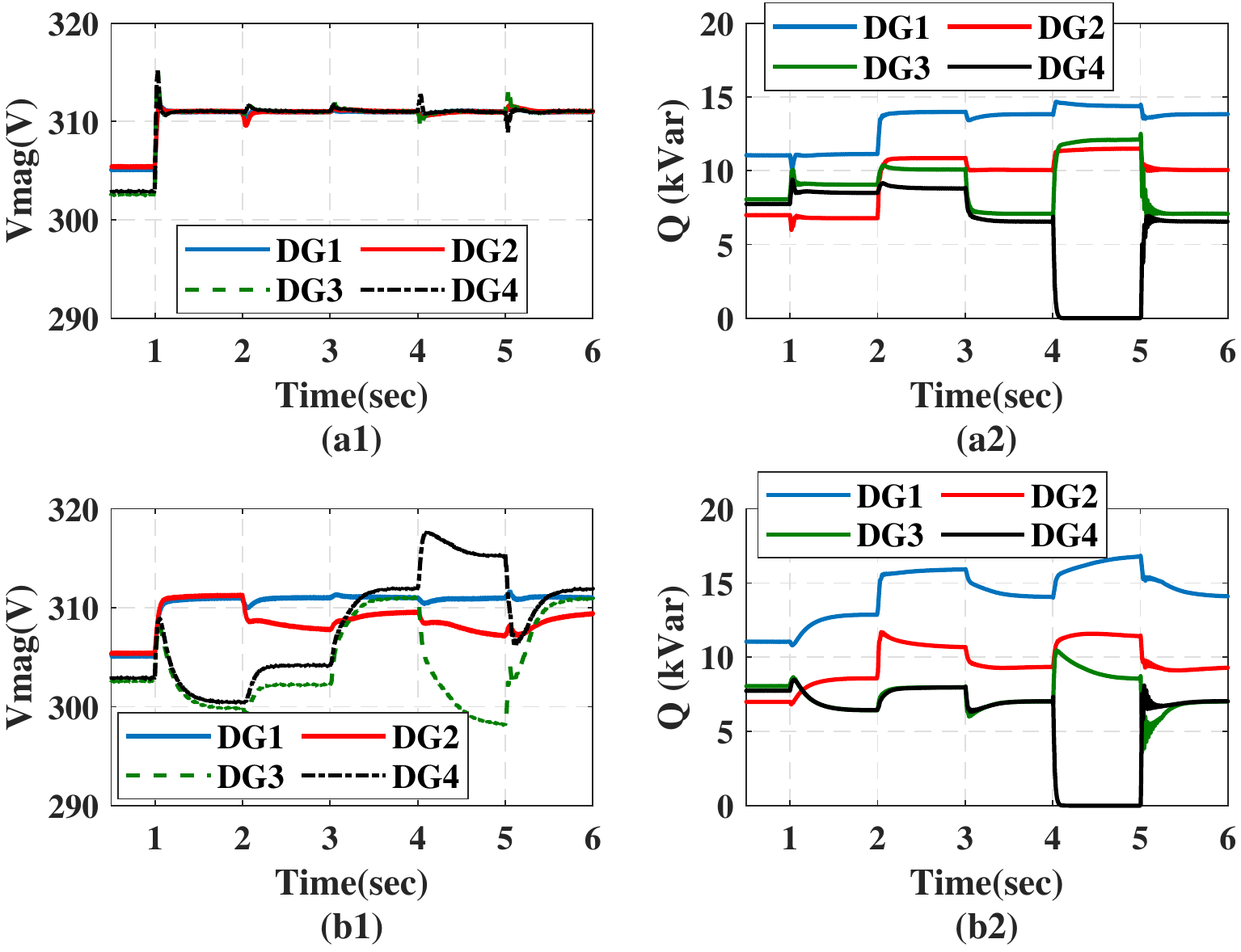}
\caption{Voltage regulation and reactive power sharing: (a) voltage regulation without reactive power sharing, (b) accurate reactive power sharing with tight voltage regulation around reference value.}\label{fig_tradeoff}
\end{figure}

To show the conflict between voltage regulation and accurate reactive power sharing and to demonstrate the effectiveness of the proposed control framework under alternative control objectives, this subsection compares the performance in the noise-containing environment (load change at 2 s and 3 s; plug-and-play at 4 s and 5 s) under the  control laws that emphasize either voltage regulation or reactive power sharing. Compared to Fig. \ref{fig_tradeoff}(a) that only guarantees exact voltage regulation without considering reactive power sharing, the results in Fig. \ref{fig_tradeoff}(b) demonstrate the feasibility to achieve the accurate reactive power sharing with tight voltage regulation around reference value.

\subsection{Scalability Test}

\begin{figure}[!t]
\centering
\includegraphics[width=2in]{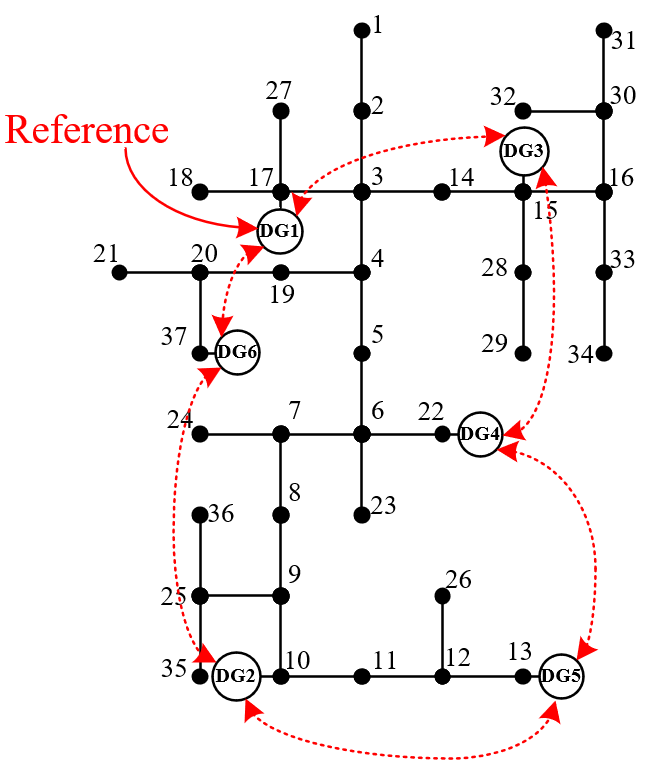}
\caption{Diagram of the modified IEEE 37-bus system.}\label{fig_IEEE37_diagram}
\end{figure}

\begin{figure}[!t]
\centering
\includegraphics[width=0.45\textwidth]{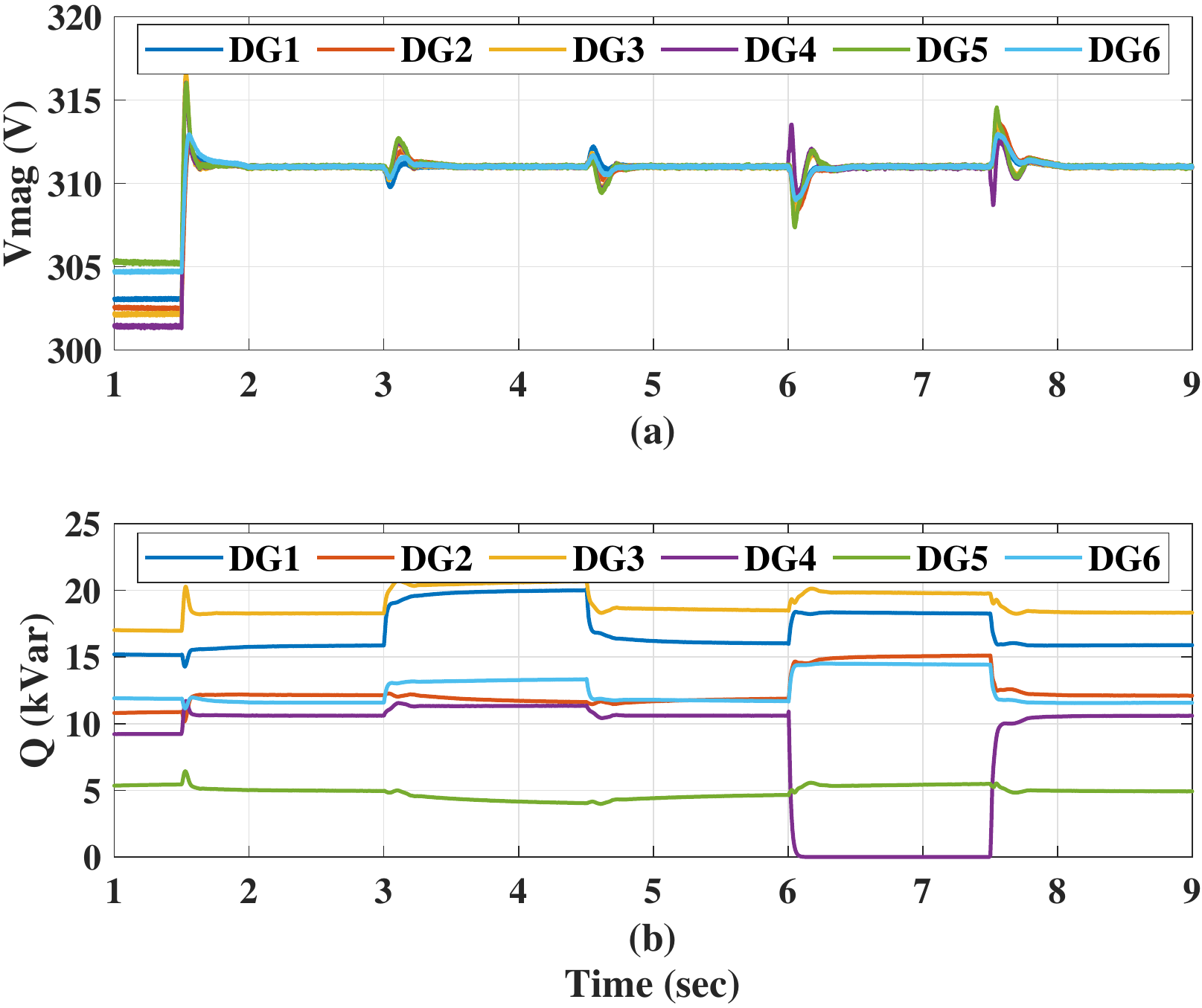}
\caption{Scalability evaluation: (a) voltage magnitude, (b) reactive power output.}\label{fig_IEEE37_Vmag}
\end{figure}

The scalability of the proposed control method is investigated in the modified IEEE 37 bus system~\cite{nourollah_combinational_2016}, as shown in Fig.~\ref{fig_IEEE37_diagram}. 
Before $t=1.5$ s, the 37-node MG system operates in the islanded model with the total loads of 122.10 kW and 70.35 kVar, and DGs are controlled under the primary control only. After $t=1.5$ s, the proposed secondary voltage control is activated. The loads of 15.75 kW and 7.88 kVar are increased and decreased on node 2 at $t=3$ s and $t=4.5$ s respectively, and DG4 is disconnected and re-connected at $t=6$ s and $t=7.5$ s respectively.

\begin{figure}[!t]
\centering
\includegraphics[width=0.45\textwidth]{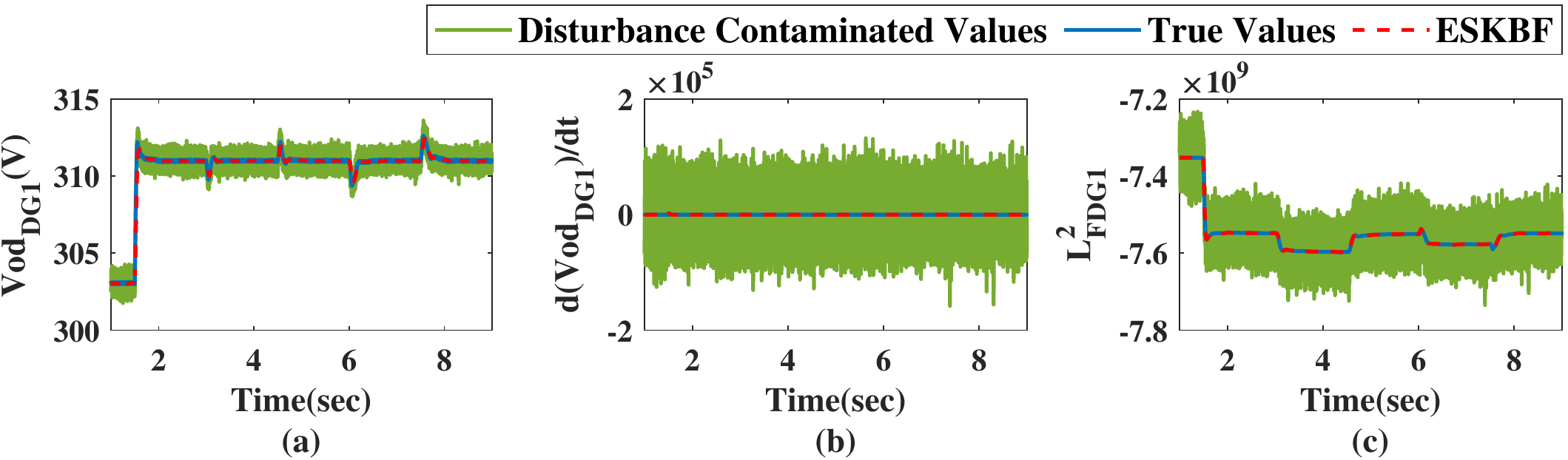}
\caption{ESKBF-based observer performance evaluation (DG1) of the scalability test: (a) $v_{odi}$, (b) $\dot{v}_{odi}$, (c) $L_{\boldsymbol{F}_i}^{2}h_i(\boldsymbol{x}_i)$.}\label{fig_IEEE37_ESKBF}
\end{figure}

As shown in Fig.~\ref{fig_IEEE37_Vmag}, by applying the proposed secondary voltage control method, the output voltages of DGs can be regulated to the reference when load change and plug-and-play occur. Moreover, the performance of ESKBF is similar to that in the 4-bus MG system, so for the sake of simplification, only the observation performance of DG1 is shown in Fig.~\ref{fig_IEEE37_ESKBF}.

\subsection{Experimental Verification}

\begin{figure}[!t]
\centering
\includegraphics[width=0.45\textwidth]{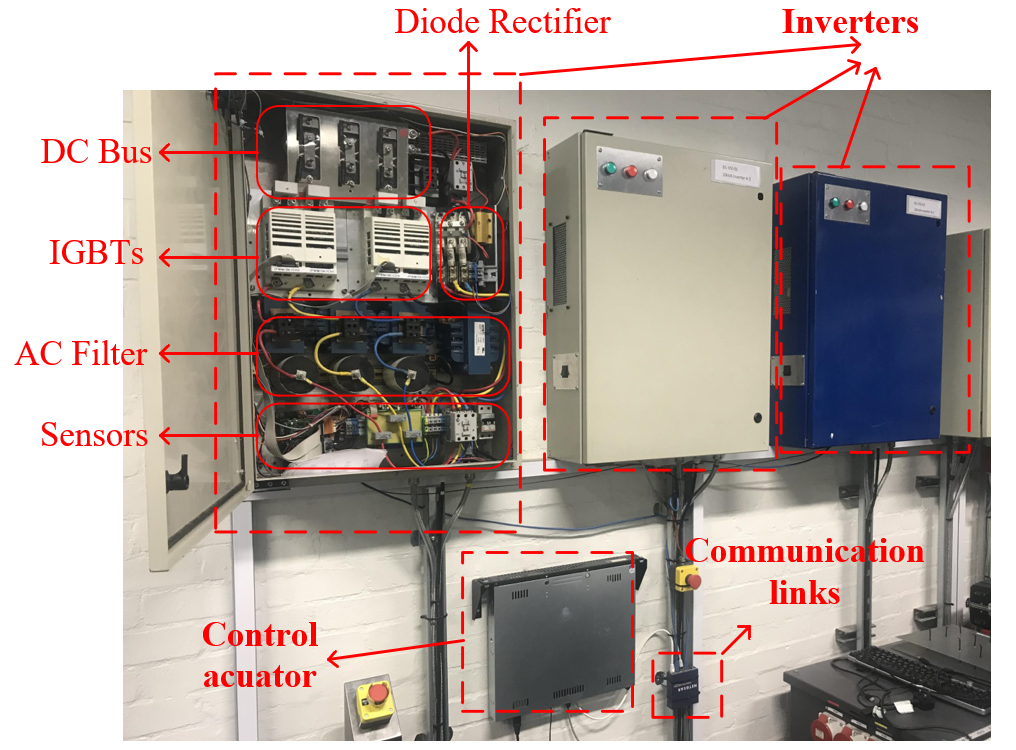}
\caption{Experimental testbed of the MG with three inverters.}\label{fig_testbed}
\end{figure}

\begin{figure}[!t]
\centering
\includegraphics[width=0.4\textwidth]{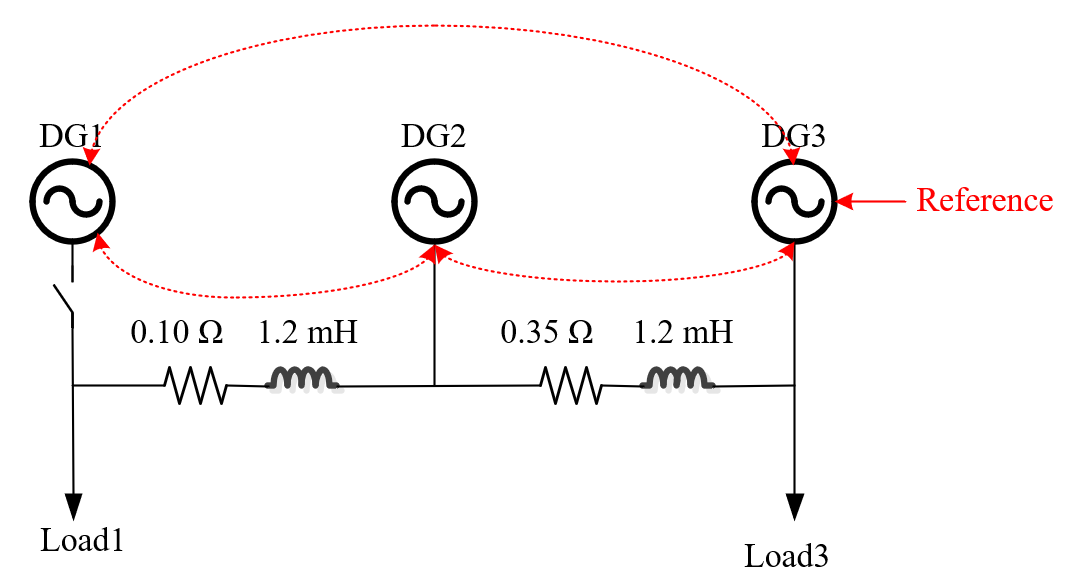}
\caption{Topology of the experimental testbed.}\label{fig_exptopo}
\end{figure}

To validate the effectiveness of the proposed control method in a practical scenario, an experimental MG testbed, as shown in Fig.~\ref{fig_testbed} with three inverters has been developed to test the control performance. The topology of the MG testbed is shown in Fig.~\ref{fig_exptopo}, and the parameters of the inverter are the same as the Table I in \cite{pogaku_modeling_2007} ($v_{ref}=381$ V).

\begin{figure}[!t]
\centering
\includegraphics[width=0.43\textwidth]{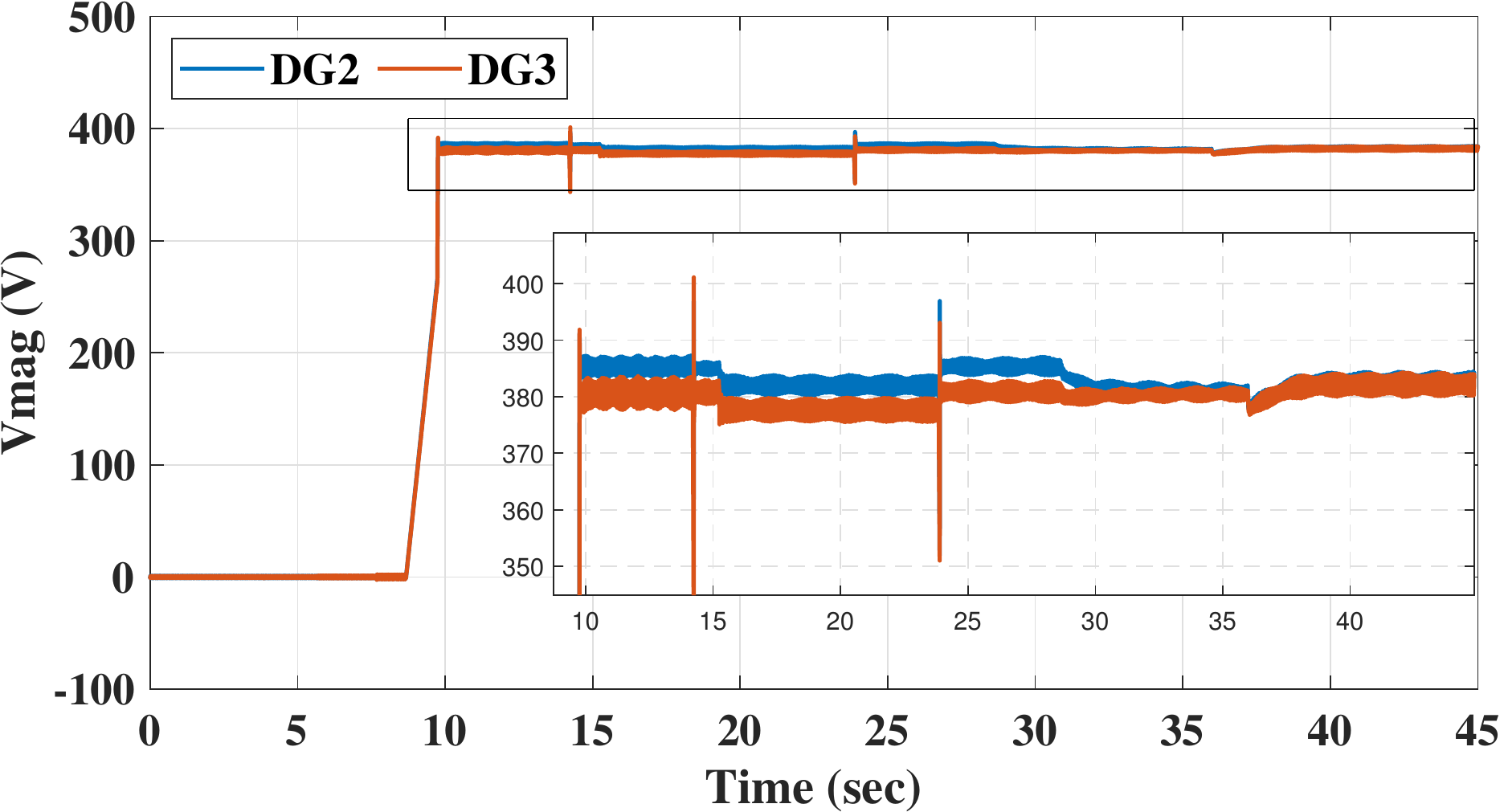}
\caption{Voltage control performance in the experimental scenario with load change.}\label{fig_exp_V}
\end{figure}

\begin{figure}[!t]
\centering
\includegraphics[width=0.45\textwidth]{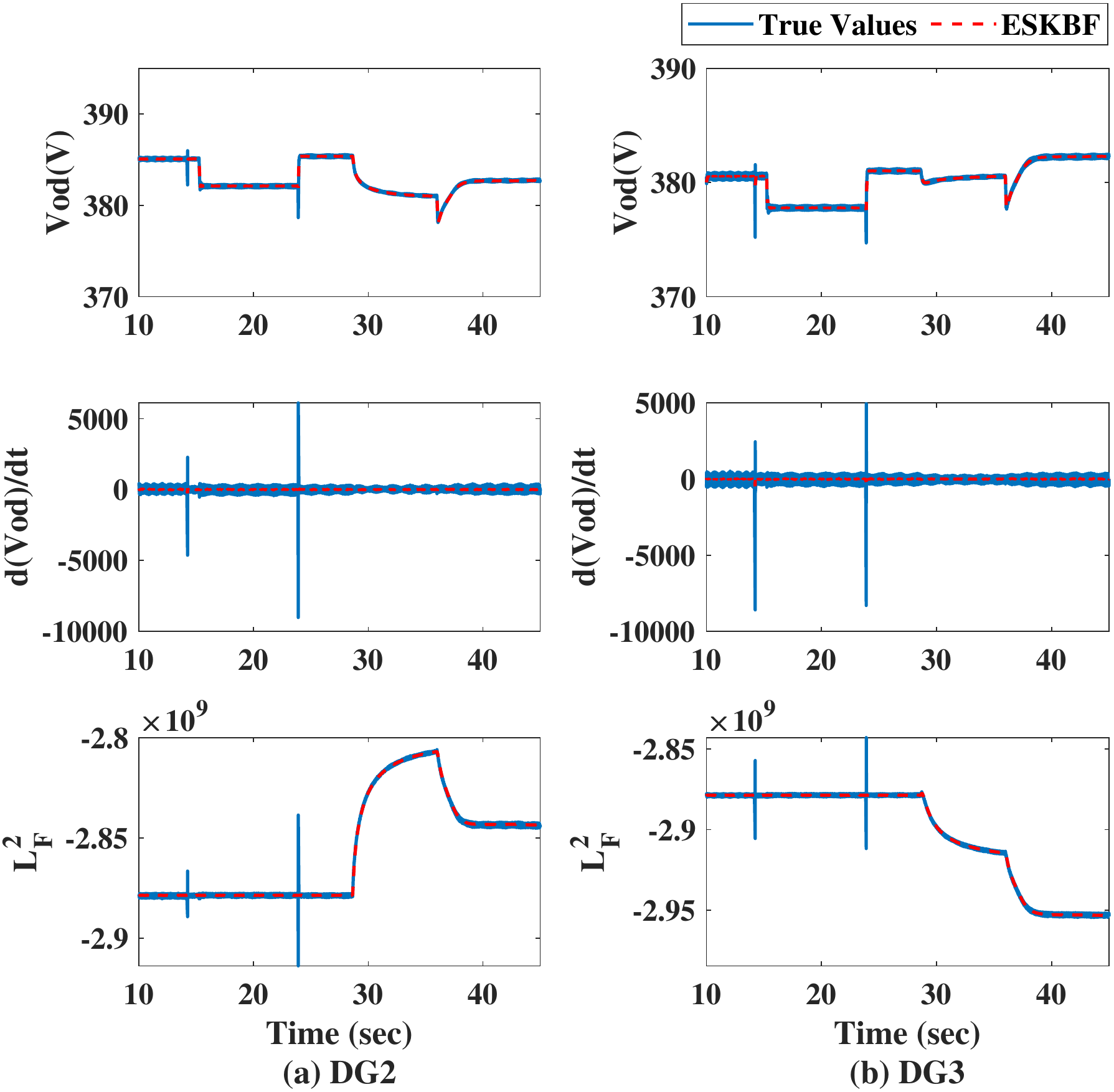}
\caption{ESKBF performance in the experimental scenario with load change.}\label{fig_exp_ESKBF}
\end{figure}

Two experimental cases are designed, including load change and plug-and-play capability test. The control performance is detailed as in Fig.~\ref{fig_exp_V} when load change occurs. Throughout the whole experiment, the active power load (Load3) connected at DG3 is 3 kW. 
%DG2 and DG3 are connected to the MG at $t=8.5$ s and $t=14$ s respectively. 
DG2 and DG3 are switched in and started up during the periods $8.5-10$ s and $14-16$ s respectively.
A capacitance ($50\ \mu F$) is connected to the grid at $t=24$ s, driving the increase of the output voltages of DG2 and DG3 due to the capacitive reactive power output. The voltages of two DGs are regulated to the reference when the proposed secondary control is activated at $t=28.5$ s. At $t=36$~s, the capacitance is disconnected from the MG, and the voltages are restored as well. The performance of ESKBF during this experiment is detailed in Fig.~\ref{fig_exp_ESKBF}, demonstrating its fast convergence tracking property.

\begin{figure}[!t]
\centering
\includegraphics[width=0.45\textwidth]{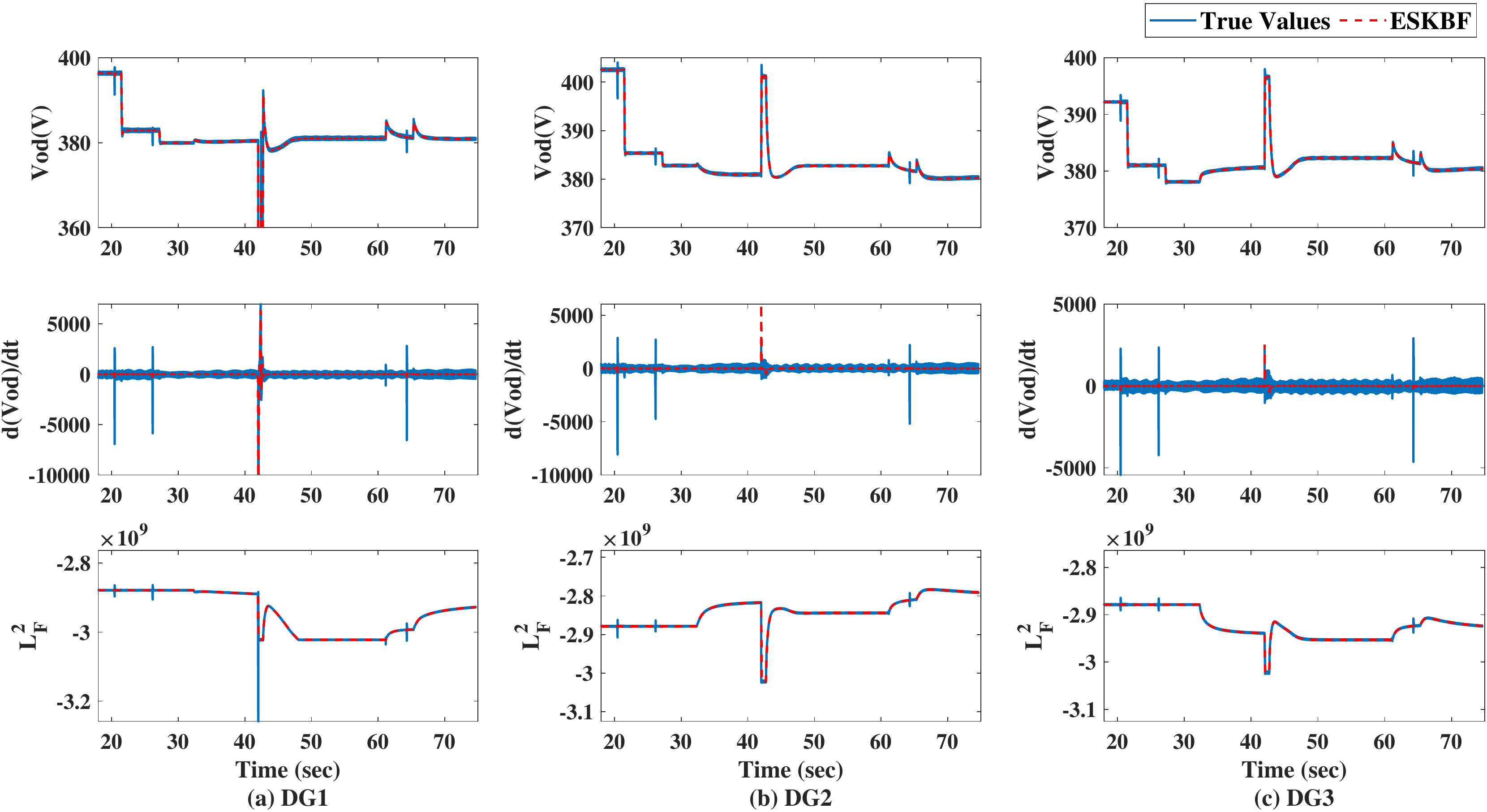}
\caption{ESKBF performance of plug-and-play capability test in the experimental scenario.}\label{fig_exp_ESKBF2}
\end{figure}

\begin{figure}[!t]
\centering
\includegraphics[width=0.45\textwidth]{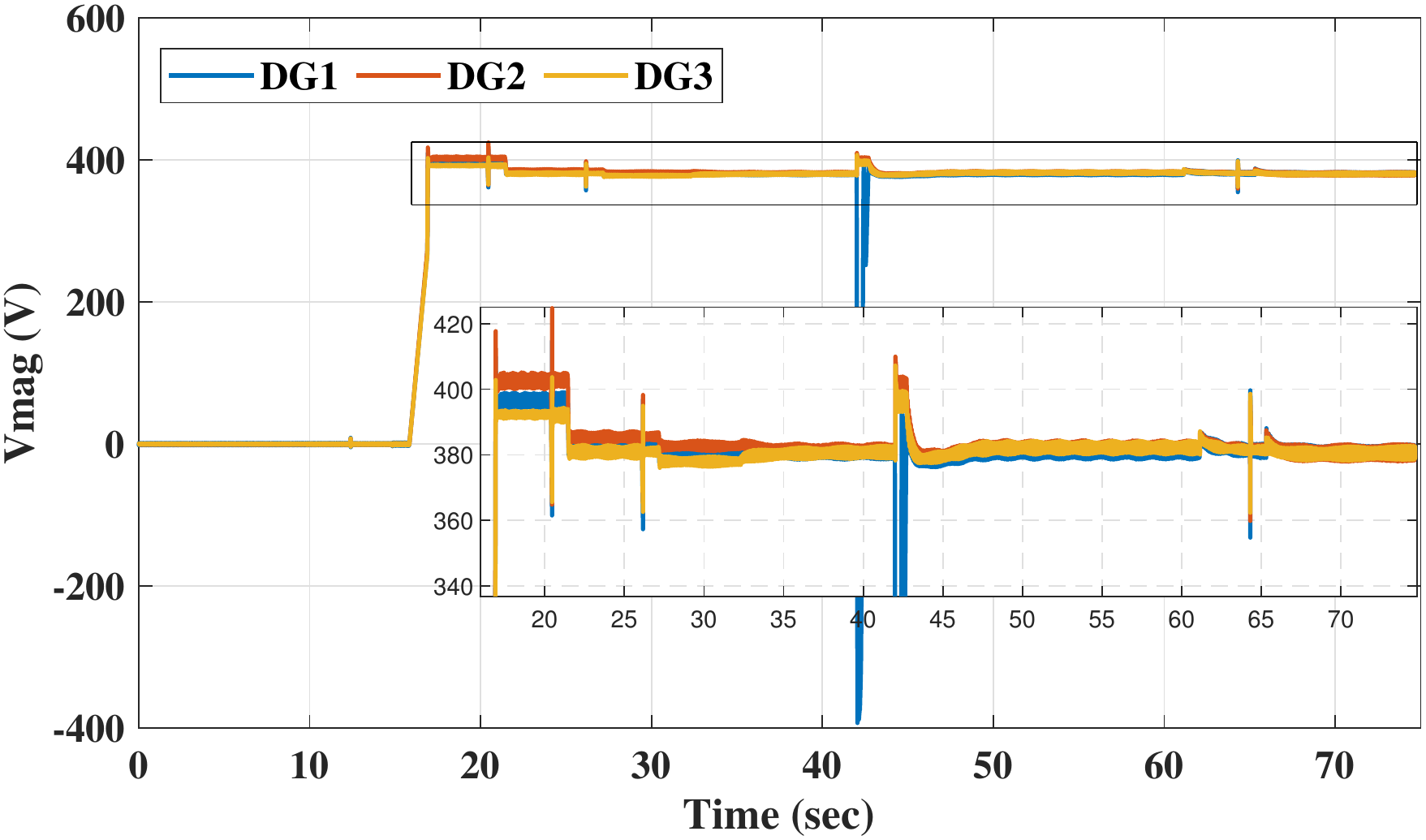}
\caption{Voltage control performance of plug-and-play capability test in the experimental scenario.}\label{fig_exp_V2}
\end{figure}

The voltage control performance of the plug-and-play capability test is shown in Fig.~\ref{fig_exp_V2}, and the corresponding ESKBF-based observer performs as in Fig.~\ref{fig_exp_ESKBF2}. 
%Three DGs are connected to the MG at $t=$16 s, 21 s, 26 s respectively. 
Three DGs are switched in and started up during the periods $16-18$ s, $20-22$ s and $26-27.5$ s respectively.
The proposed secondary control is activated at $t=32.5$ s, when the voltages of three DGs start to synchronized to the reference. The DG1 is disconnected at $t=42$ s, after a very short transient, the voltages are correctly restored. The re-connection of DG1 occurs from $t=61$ s to $t=64.5$ s, including the re-connection of LC filter and the re-activation of DG1 inner control loops. %\footnotemark[1]. 
The corresponding voltage restoration illustrates the effectiveness of the proposed secondary voltage control when the plug-in process occurs.
%\footnotetext[1]{The modified frequency control for plug-and-play operation in the practice is detailed in the Appendix.}

\section{Conclusion}\label{section_conclusion}
This paper proposes a distributed secondary voltage control method for the resilient operation of an islanded microgrid, where each inverter-based DG is modeled as an agent and the MG is seen as a multi-agent system by using the sparse communication network. Firstly, a Kalman-Bucy filter based extended state observer is designed to overcome the disturbance caused by three different sources, namely measurement noise, parameter perturbation and immeasurable variables. The ESKBF locally observes the control variables that are related to MG secondary voltage control. Then, based on the locally observed states of the DGs and distributed communication network, the proposed FTSM control employs nonlinear terminal attractors to enhance the consensus convergence rate of the system. Finally, the effectiveness of the proposed control method is illustrated by simulation and experimental studies.

The proposed secondary controller has been proved to be effective when handling different kinds of disturbance in the MG system, but there may exist intentional and external interference such as cyber attack and communication fault. Thus, further investigation is required to expand this work to tackle the problem of cybersecurity-oriented MG control design.
% or

%{\color{blue} \appendix[Frequency Control Loop for Plug-and-Play]\label{appendix}
%\begin{figure}[htb]
%\centering
%\includegraphics[width=0.45\textwidth]{PLL_syn.png}
%\caption{Fequency control loop}\label{fig:PLL}
%\end{figure}

%The frequency control loop, as shown in Fig. \ref{fig:PLL} is applied in the droop-based inverters. If there is no plug-and-play, the frequency control loop is the same as a traditional grid-forming inverter, and this loop is specially used for phase synchronization when plug-and-play occurs. In the practice, the LC filter is connected to the grid before the inverter itself. Then, the phase-locked loop (PLL) can obtain the corresponding phase of the inverter output capacitance to form synchronization. The variable \textit{enable} represents the state of DG plugged out or in (0: out; 1: in). When one DG plugged in, \textit{enable} change from 0 to 1, and the frequency control can do the integral of the frequency with the initial phase that comes from the PLL.
\ifCLASSOPTIONcaptionsoff
  \newpage
\fi



%\bibliographystyle{IEEEtran}

\bibliography{references} 
\bibliographystyle{IEEEtran}

\vspace{-30pt}
\begin{IEEEbiography}[{\includegraphics[width=1in,height=1.25in,clip,keepaspectratio]{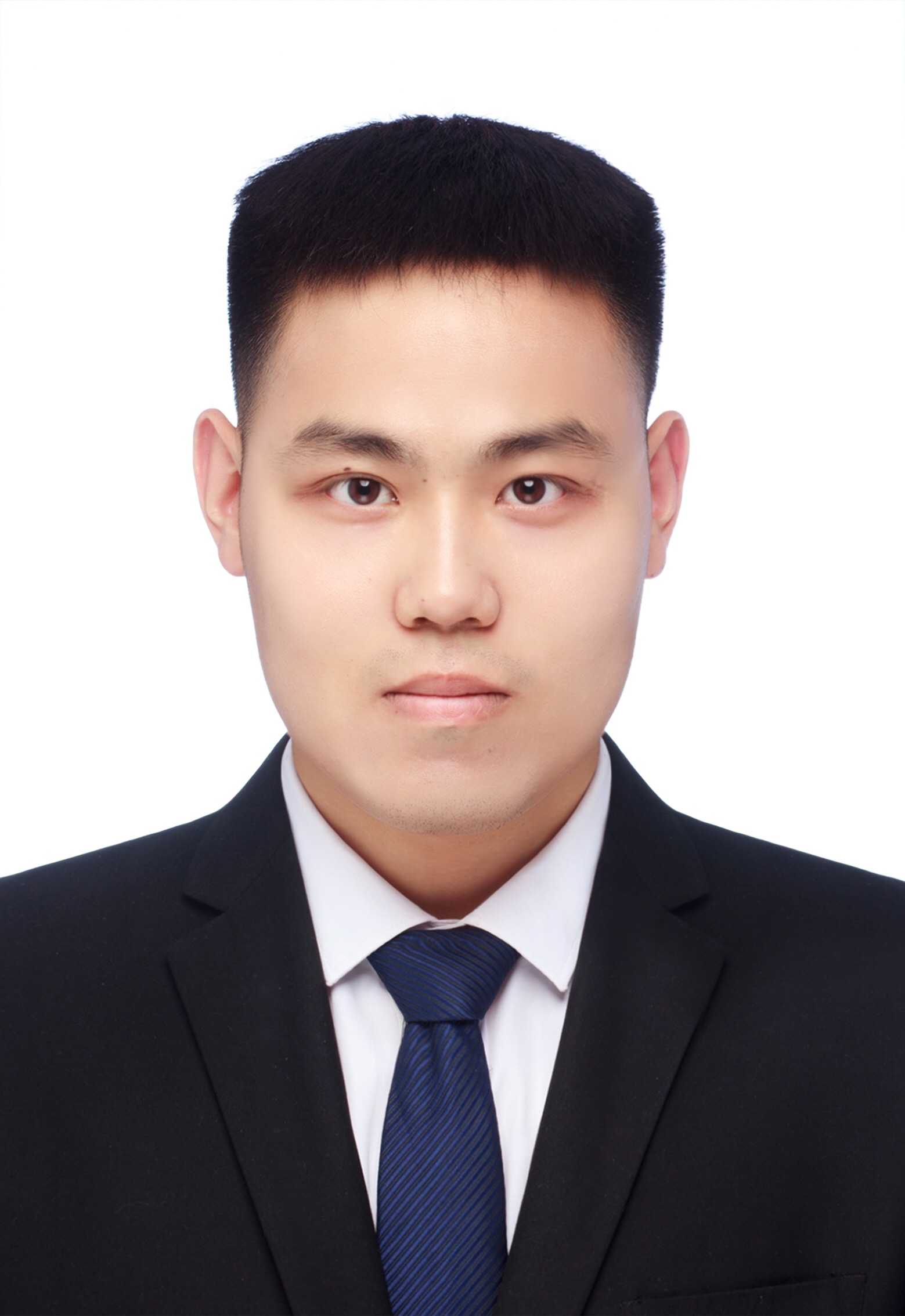}}]{Pudong~Ge}
(S'18) received the M.Sc degree in electrical engineering from Southeast University, Nanjing, China, in 2019, and he is currently a Ph.D student at Department of Electrical and Electronic Engineering, Imperial College London. His current research focuses on the distributed control of cyber-physical coupling microgrids, and cyber-resilient energy system operation and control.
\end{IEEEbiography}

\begin{IEEEbiography}[{\includegraphics[width=1in,height=1.25in,clip,keepaspectratio]{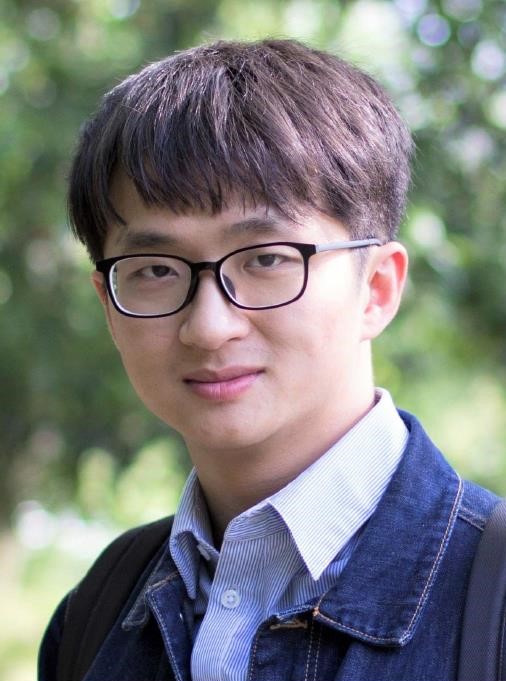}}]{Yue~Zhu}
received the B.Eng and M.Sc degrees in electrical engineering, Zhejiang University, Hangzhou, China, in 2016 and 2019 respectively. He is currently a Ph.D student at Department of Electrical and Electronic Engineering, Imperial College London. His present research focuses on impedance-based stability analysis of power systems, and the noise evaluation for impedance measurement.
\end{IEEEbiography}

\begin{IEEEbiography}[{\includegraphics[width=1in,height=1.25in,clip,keepaspectratio]{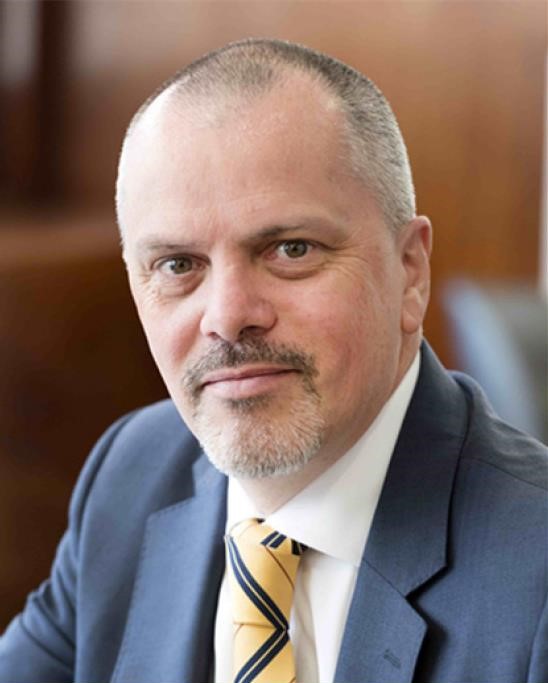}}]{Tim~C.~Green}
(M'89-SM'02-F'19) received a B.Sc. (Eng) (first class honours) from Imperial College London, UK in 1986 and a Ph.D. from Heriot-Watt University, Edinburgh, UK in 1990. He is a Professor of Electrical Power Engineering at Imperial College London, and Director of the Energy Futures Lab with a role fostering interdisciplinary energy research. His research interest is in using the flexibility of power electronics to accommodate new generation patterns and new forms of load, such as EV charging, as part of the emerging smart grid. In HVDC he has contributed converter designs that reduce losses while also providing control functions assist AC system integration. In distribution systems, he has pioneered the use of soft open points and the study of stability of grid connected inverters. Prof. Green is a Chartered Engineering the UK and a Fellow of the Royal Academy of Engineering.
\end{IEEEbiography}
\vspace{-450pt}
\begin{IEEEbiography}[{\includegraphics[width=1in,height=1.25in,clip,keepaspectratio]{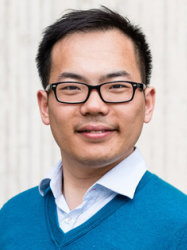}}]{Fei~Teng}
(M'15) received the BEng in Electrical Engineering from Beihang University, China, in 2009, and Ph.D. degree in Electrical Engineering from Imperial College London, U.K., in 2015. Currently, he is a Lecturer in the Department of Electrical and Electronic Engineering, Imperial College London, U.K. His research focus on the efficient and resilient operation of future cyber-physical power system.
\end{IEEEbiography}
\end{document}